\documentclass[12pt]{article}
\usepackage{natbib}
\usepackage{amsmath,amssymb,amsfonts}
\usepackage{graphicx}
\usepackage{textcomp}
\usepackage[svgnames]{xcolor}
\usepackage{colortbl}

\usepackage{bbding}

\newcommand\YES{\scriptsize\Checkmark}

\usepackage[inline]{enumitem}
\usepackage{placeins}

\usepackage[framemethod=tikz]{mdframed}
\mdfdefinestyle{mpdframe}{
    frametitlebackgroundcolor   =black!15,
    frametitlerule              =true,
    roundcorner                 =3pt,
    middlelinewidth             =0.8pt,
    innermargin                 =0.1cm,
    outermargin                 =0.1cm,
    innerleftmargin             =0.1cm,
    innerrightmargin            =0.1cm,
    innertopmargin              =0.1cm,
    innerbottommargin           =0.1cm
}

\usepackage{caption,subcaption}
\usepackage{pgfplots,pgfplotstable}
\pgfplotsset{compat=1.14} 
\usetikzlibrary{pgfplots.groupplots}
\usetikzlibrary{external}
\usepackage{fontawesome}

\usepackage{array,makecell,tabularx}
\usepackage{multirow}
\usepackage{longtable}
\usepackage{booktabs}
\usepackage{rotating}
\newcommand\SW[1]{\begin{turn}{90}#1\end{turn}}
\usepackage{csvsimple-legacy}

\usepackage{xspace}
\usepackage[autolanguage]{numprint}
\AtBeginDocument{\npthousandsep{\,}}
\usepackage{fp}

\newcommand*\np[2][z]{
\ifx z#1%
$\numprint{#2}$%
\else%
$\numprint[#1]{#2}$%
\fi\xspace%
}

\newcommand{\ShowAbsoluteNumber}[1]{%
\ifnum #1<10%
{\hspace*{0pt}#1}%
\else%
\ifnum #1<100%
{\hspace*{0pt}#1}%
\else%
\ifnum #1<1000%
{\hspace*{0pt}#1}%
\else%
{\numprint{#1}}%
\fi%
\fi%
\fi%
}

\newcommand{\ShowPercentage}[2]{%
\FPeval\percentage{round(#1/#2*100,0)}%
\FPeval\percentageOneDecimal{round(#1/#2*100,1)}%
\ifnum \percentage=0%
{\np[\%]{\FPprint{percentageOneDecimal}}}%
\else%
\ifnum \percentage<10%
{\np[\%]{\FPprint{percentageOneDecimal}}}%
\else%
{\np[\%]{\FPprint{percentageOneDecimal}}}%
\fi%
\fi%
\xspace
}
\newlength\BARSIZE  
\newcommand{\inlinechart}[2]{%
\FPeval{\BLACKBARSIZE}{#1/#2}\textcolor{black!80}{\rule{\BLACKBARSIZE\BARSIZE}{1.6ex}}%
\FPeval{\BLACKBARSIZE}{1 - (#1/#2)}\textcolor{black!10}{\rule{\BLACKBARSIZE\BARSIZE}{1.6ex}}%
}

\newcommand*\percent[3][v]{%
\ifx b#1
    (\np{#2} / \ShowPercentage{#2}{#3})\else%
\ifx q#1%
    \np{#2}/\np{#3}(\ShowPercentage{#2}{#3})\else%
\ifx v#1%
    \ShowPercentage{#2}{#3}\else%
\ifx p#1%
    \np{#2}(\ShowPercentage{#2}{#3})\else%
\ifx c#1%
    \inlinechart{#2}{#3}%
\else%
    \np{#2}%
    \ifx r#1%
        /\np{#3}%
    \fi%
    \hspace*{0.5ex}(\ShowPercentage{#2}{#3}) %
    \inlinechart{#2}{#3}%
    \xspace
\fi\fi\fi\fi\fi%
}

\usepackage{solidity}

\newcommand\Cpp{{C\nolinebreak[4]\hspace{-.05em}\raisebox{.2ex}{\scriptsize\bfseries ++}}}
\newcommand\OP[1]{\texttt{\small#1}}

\newcommand\ITEM[1]{\par\smallskip\noindent\textit{#1}}
\newcommand\TODO[1]{%
  \ifx&#1&%
    \textcolor{red}{TODO!} 
  \else
    \begin{flushleft}
      \color{red}%
      TODO: #1%
    \end{flushleft}
  \fi
  \typeout{*** TODO: #1}%
}

\AtBeginDocument{%
}

\usepackage{siunitx}

\catcode`_=\active
\catcode`_=8

\newcommand\SB[1]{\texttt{#1}}
\newcommand\TAG[2][\relax]{%
  \medskip\par\noindent\SB{#2}%
  \ifx\relax#1\else
    \ (#1)%
  \fi
  .
}

\ExplSyntaxOn
\cs_set_eq:NN \explower \text_lowercase:n
\ExplSyntaxOff

\usepackage{swc}

\definecolor{one}{HTML}{4F81BD}
\definecolor{two}{HTML}{C0504D}
\definecolor{three}{HTML}{9BBB59}
\definecolor{more}{HTML}{9F4C7C}

\definecolor{mythril}{HTML}{D6D2D2}
\definecolor{conkas}{HTML}{F1E4F3}
\definecolor{osiris}{HTML}{F4BBD3}
\definecolor{madmax}{HTML}{F686BD}
\definecolor{ethainter}{HTML}{FE5D9F}
\definecolor{securify}{HTML}{9B7874}
\definecolor{teether}{HTML}{666370}
\definecolor{pakala}{HTML}{1C1F33}
\definecolor{ethor}{HTML}{A6B1E1}
\definecolor{maian}{HTML}{B4869F}

\definecolor{bad}{HTML}{FAAAAD}
\definecolor{mid_bad}{HTML}{FCE6E9}
\definecolor{mid_good}{HTML}{BDE2C8}
\definecolor{good}{HTML}{63BE7B}

\definecolor{midblue}{rgb}{0.70,0.78,0.93}
\definecolor{lightblue}{rgb}{0.85,0.95,0.99}
\definecolor{lightgreen}{rgb}{0.74,0.89,0.78}
\definecolor{lightgray}{HTML}{F3F3F3}
\definecolor{greendark}{HTML}{6EC284}
\definecolor{greenmedium}{HTML}{7AC78E}
\definecolor{greenlight}{HTML}{BCE2C8}
\definecolor{redlight}{HTML}{FCE6E9}
\definecolor{redmedium}{HTML}{FBB6B8}
\definecolor{reddark}{HTML}{F8696B}

\usepackage{orcidlink}

\begin{document}

\title{Evolution of Automated Weakness Detection in Ethereum Bytecode: a Comprehensive Study}

\author{%
  \begin{tabular}{@{}cc@{}}
  \begin{tabular}[t]{@{}c@{}}
    Monika di Angelo \orcidlink{0000-0002-4217-4530}\\
    \normalsize TU Wien, Vienna, Austria\\
    \normalsize monika.di.angelo@tuwien.ac.at
  \end{tabular}%
  &
  \begin{tabular}[t]{@{}c@{}}
    Thomas Durieux \orcidlink{0000-0002-1996-6134}\\
    \normalsize TU Delft, Delft, Netherlands\\
    \normalsize thomas@durieux.me
  \end{tabular}%
  \\[1.5cm]
  \begin{tabular}[t]{@{}c@{}}
    Jo\~{a}o~F.~Ferreira \orcidlink{0000-0002-6612-9013}\\
    \normalsize INESC-ID and Instituto Superior T\'ecnico\\
    \normalsize University of Lisbon, Lisbon, Portugal\\
    \normalsize joao@joaoff.com
  \end{tabular}%
  &
  \begin{tabular}[t]{@{}c@{}}
    Gernot Salzer \orcidlink{0000-0002-8950-1551}\\
    \normalsize TU Wien, Vienna, Austria\\
    \normalsize gernot.salzer@tuwien.ac.at
  \end{tabular}
  \end{tabular}
}
\date{31 October 2023}
\maketitle

\newpage
\begin{abstract}
Blockchain programs (also known as smart contracts) manage valuable assets like cryptocurrencies and tokens, and implement protocols in domains like decentralized finance (DeFi) and supply-chain management. These types of applications require a high level of security that is hard to achieve due to the transparency of public blockchains. Numerous tools support developers and auditors in the task of detecting weaknesses. As a young technology, blockchains and utilities evolve fast, making it challenging for tools and developers to keep up with the pace.

In this work, we study the robustness of code analysis tools and the evolution of weakness detection on a dataset representing six years of blockchain activity. We focus on Ethereum as the crypto ecosystem with the largest number of developers and deployed programs. We investigate the behavior of single tools as well as the agreement of several tools addressing similar weaknesses.

Our study is the first that is based on the entire body of deployed bytecode on Ethereum's main chain. We achieve this coverage by considering bytecodes as equivalent if they share the same skeleton. The skeleton of a bytecode is obtained by omitting functionally irrelevant parts. This reduces the 48 million contracts deployed on Ethereum up to January 2022 to 248\,328 contracts with distinct skeletons. For bulk execution, we utilize the open-source framework SmartBugs that facilitates the analysis of Solidity smart contracts, and enhance it to accept also bytecode as the only input. Moreover, we integrate six further tools for bytecode analysis. The execution of the 12 tools included in our study on the dataset took 30 CPU years. While the tools report a total of 1\,307\,486 potential weaknesses, we observe a decrease in reported weaknesses over time, as well as a degradation of tools to varying degrees.
\end{abstract}
\newpage
\section{Introduction}
\label{sec:introduction}

\emph{Smart contracts} are event-driven programs running on the nodes of decentralized networks known as \emph{blockchains}.
Specific transactions, once included in the blockchain, trigger the execution of these blockchain programs.
Every node executes the code locally within a virtual machine and updates its state of the blockchain.
The computations are deterministic, ensuring that all nodes arrive at the same state.
The flexibility of smart contracts and the unique properties of blockchains, most notably decentralization and immutability, gave rise to innovative applications in areas like decentralized finance and supply chain management.
Their potential has led to ecosystems with large numbers of start-ups and market caps of hundreds of billions of USD.

Against this background, errors in smart contracts can lead, and have led, to costly disruptions and losses.
Early on, academia and industry focused on methods and tools for developing \emph{secure} smart contracts.
In a survey on automated vulnerability detection conducted in mid-2021, \cite{Rameder2022} identified 140 tools for Ethereum, the major smart contract platform.
The sheer number makes it hard to decide which tools may be suited for the task at hand, and calls for regular tool evaluations and comparisons.

In this paper, we present a comprehensive evaluation of 12 tools for vulnerability detection on the Ethereum main chain. The goal of this study is to analyze how typical tools behave within the Ethereum ecosystem. We focus on the evolution of tools and their findings to identify common patterns and trends. The results of our study can be utilized to inform developers about the state of the art in automated vulnerability detection and to guide researchers in the development of new tools. Additionally, it provides an overview of the reliability of the selected tools and whether they should be included in future studies.

Our work differs from previous studies in several aspects. First, we analyze the \emph{temporal evolution of weakness detection}, focusing on the robustness of tools over time (rather than assessing their detection capabilities against a set of contracts).

Second, we aim at a \emph{complete coverage of the Ethereum main chain}, which is a formidable endeavor in light of 48 million deployments of smart contracts (up to Jan 2022).
This enables us to investigate the evolution of weakness detection over a period of more than six years.
We select one contract per skeleton of bytecode (cf.\ \autoref{sec:data}), which reduces the number of objects to analyze to \np{248328}.

Third, we concentrate on the \emph{runtime bytecode} as input to the tools.
Surveys usually evaluate tools on benchmarks of Solidity source code (cf.\ \autoref{sec:related}), and hence omit tools analyzing bytecode only.
Moreover, for many contracts on the blockchain, the source code is not available.
By choosing runtime bytecode as the least denominator, we can include tools rarely considered, and we are able to analyze all smart contracts deployed so far.

Finally, to perform our study, we extend SmartBugs~\citep{ferreira2020smartbugs}, a framework for executing analysis tools in a unified manner.
Integrating new tools into the framework makes them available for future evaluations by others.

With 12 tools, 15 weakness classes, \np{248328} runtime bytecodes of smart contracts and an execution time of 30 years, our evaluation is more comprehensive than previous studies.
The large number of samples as well as the method of selection allows us to add a unique temporal perspective.
In summary, the contributions of this paper are:
\begin{itemize}
\item A method for selecting a feasible number of smart contracts that are representative of \np[M]{48} blockchain programs deployed on Ethereum in the course of six years.
\item A public dataset of \np{248328} smart contracts that may serve as the basis of further evaluations.%
  \footnote{Available at \url{https://github.com/gsalzer/skelcodes}}
\item An extension of the framework SmartBugs to include 12 tools for vulnerability detection with bytecode-only input.%
  \footnote{Available at \url{https://github.com/smartbugs/smartbugs}}
\item Methods for analyzing and visualizing the temporal evolution of tool results and the overlap between tools.
\item A portrait of the evolution of tool behavior and weakness detection on six years of blockchain activity.
\end{itemize} 

\newpage
\section{Study Design}
\label{sec:design}

Our study aims to provide a comprehensive perspective on the evolution of weaknesses in smart contracts, as detected by automated analysis tools.
We focus on the Ethereum blockchain, which is the major platform for smart contracts in terms of the number of applications, market cap, attacks, and countermeasures. 
To address the research questions outlined below, we proceed as follows.

\ITEM{Collecting the Contract Data.}
The study period covers the 14 million blocks from Ethereum's start on 30~July 2015 up to 13~Jan 2022.
During this period, there were 48 million contract deployments. 
Analyzing the contracts in their entirety is not only infeasible, but wastes resources and introduces biases, as the deployments range from one-of-a-kind contracts to code deployed hundreds of thousands of times.
In \autoref{sec:data}, we introduce the \emph{skeleton} of contracts.
By grouping contracts with identical skeletons and selecting only one contract per group, we capture the diversity of the Ethereum ecosystem by analyzing just \np{248328} contracts.
We assess various properties of the data needed later on. 

\ITEM{Selecting the Analysis Tools.}
Performing a large-scale analysis on smart contracts that, in general, are only available as bytecode, restricts the number of applicable tools.
In \autoref{sec:tools}, we specify the selection criteria and apply them to the \np{140} tools identified by \cite{Rameder2022}.
We describe the 12 tools that remain regarding engineering aspects and their approach to contract analysis.

\ITEM{Analyzing the contracts.}
To execute 12 tools with diverse requirements and I/O formats on \np{248328} contracts, we select the execution framework~\emph{SmartBugs}.
Initially, it contained only half of the needed tools and required Solidity source code as input.
We extended SmartBugs to accept bytecode as well and added the other six tools.
After a cumulative execution time of 30 years, we obtain three million records, each specifying the result of running a single tool on a specific bytecode.
We refer to \autoref{sec:execution} for the details regarding the execution framework.

\ITEM{Mapping the Weaknesses to a Common Taxonomy.}
The execution data allows us to analyze the detection results per tool.
To facilitate the comparison and aggregation of results from multiple tools, we map the tool findings to a common taxonomy that is described in \autoref{sec:weaknesses}.

\ITEM{}%
Based on the results of running the analyzers on the bytecodes, we address the following research questions.

\ITEM{\bfseries RQ1 Abstraction.} \emph{How well are skeletons suited as an abstraction of functionally similar bytecode in the context of weakness analysis?}
In \autoref{sec:rq1}, we investigate whether and how the weakness analysis of bytecodes with the same skeleton differs.

\ITEM{\bfseries RQ2 Weakness Detection.} \emph{Which trends in the weakness reports of analysis tools can be identified for the contracts on Ethereum's main chain?}
In \autoref{sec:rq2}, we are interested in the evolution of types and numbers of weaknesses reported for the deployments up to early 2022.

\ITEM{\bfseries RQ3 Tool Quality.} \emph{How do analysis tools change their behavior in a weakness analysis with bytecode input?}
In \autoref{sec:rq3}, we investigate the evolution of tool quality with respect to maintenance aspects, execution time, errors, and failures. 

We do not assess the individual performance of the tools, like the rates of true/false positives/negatives, as there is no ground truth that is sufficiently large, consistent and balanced for a conclusive evaluation~\citep{MdAGS2023GT}.

\ITEM{\bfseries RQ4 Overlap Analysis.} \emph{To which extent do the tools agree when addressing similar weaknesses?}
In \autoref{sec:rq4}, we determine the overlap of tools for weaknesses that can be mapped to the same class of the SWC registry.

\ITEM{Discussion.}
In \autoref{sec:discussion}, we combine the results of our research questions.
For two specific SWC classes, we investigate how the agreement of the tools evolve over time, and provide an explanation.
Moreover, we consider the limitations of our study.

\ITEM{Related work.}
\autoref{sec:related} gives an overview of studies similar to ours and highlights the differences.
Moreover, it compares two execution frameworks and justifies our decision for using SmartBugs.
\newpage
\section{Contract Data}
\label{sec:data}

This section describes the collection and preparation of the contract data that we are going to analyze.
Moreover, we assess some characteristics of the data that will be needed to interpret the results of the analysis tools.
We start by clarifying some concepts specific to Ethereum.

\subsection{Creation of Contracts}
\label{sec:creation}

\ITEM{Deployment vs.\ runtime code.}
To deploy a contract on an Ethereum chain, an external user submits a create transaction, or the Ethereum Virtual Machine (EVM) executes a create operation.
The transaction/operation includes the \emph{deployment code}.
The code consists of an active part, $D$, which typically initializes the environment for the new contract.
At its end, $D$ returns the pointer to a memory area with the actual \emph{runtime code}, which the EVM then stores at the address of the new contract.
The deployment code is free to assemble the runtime code arbitrarily, but typically just copies code following~$D$ in the deployment code.

\ITEM{Source code vs.\ bytecode.}
The majority of Ethereum contracts are written in Solidity, a programming language inspired by \Cpp.
The so-called constructor and any global initializations compile to the active part of the deployment code, $D$, whereas all other parts of the source file compile to the runtime code proper,~$R$, which is appended to $D$.
After~$R$, the compiler appends \emph{meta-data}, $M$, which contains a hash identifying the original source code and version information.
Changing any character in the Solidity file, including comments and the newline encoding, alters $M$ and leads to superficially different deployment and runtime codes.

To illustrate the role of the different forms of code, consider the program in \autoref{fig:Solidity program}.
It shows the Solidity code of a contract \texttt{C1} that deploys a contract of type \texttt{C2} as part of its own deployment.
At runtime, each call to function \texttt{f} deploys a contract of type~\texttt{C3}.
From this source code, the compiler generates a bytecode of the form \(D_1R_1D_3R_3M_3M_1D_2R_2M_2\).
During deployment, $D_1$ creates the contract~\texttt{C2} by executing $D_2R_2M_2$, with the runtime code $R_2M_2$ getting stored at the address of \texttt{C2}.
Then $D_1$ returns $R_1D_3R_3M_3M_1$ as runtime code, which is stored at the address of~\texttt{C1}.
Later, when the function~\texttt{f} of contract~\texttt{C1} is called, $D_3R_3M_3$ gets executed and creates a new contract with runtime code $R_3M_3$.
Note the multiple occurrences of meta-data and their proliferation during deployment and runtime.

\begin{solidity}[float,basicstyle=\footnotesize\ttfamily,caption=Solidity contract creating contracts during deployment (\texttt{C2}) as well as during runtime (\texttt{C3}). The definitions of \texttt{C2} and \texttt{C3} have been omitted.,label=fig:Solidity program,numbers=none]
contract C1 {
    C2 c = new C2();
    function f() public returns(C3) {
        return new C3();
}   }
\end{solidity}

\ITEM{Skeletons.}
The skeleton of a contract is obtained by removing meta-data, the arguments of \OP{PUSH} operations, constructor arguments, and trailing zeros.
The rationale is to remove parts that contribute little to the functionality of the contract, with the aim to equate contracts with the same skeleton.

\ITEM{Code family.}
A \emph{family of codes} is a collection of runtime codes with the same skeleton.

\subsection{Data Collection}

We strive for a complete coverage of Ethereum's main chain.
Therefore, we collect the runtime codes of all contracts (including the self-destructed ones) that were successfully deployed up to block 14\,M (13~Jan 2022).%
\footnote{We used an OpenEthereum client, \url{https://github.com/openethereum/openethereum}.}
\autoref{tab:deployments} gives an overview of the deployment activities.
\begin{table}
  \caption{Deployments up to Block 14\,M (Jan 2022)}
  \label{tab:deployments}
  \centering
  \begin{tabular}{@{}lr@{}}
    \toprule
    Metric & \# Contracts \\
    \midrule
    Deployments & \np{48262411} \\
    Distinct deployment codes & \np{2206793} \\
    Distinct runtime codes & \np{514893}\\
    \dots\ without meta-data & \np{364599} \\
    \dots\ without \OP{PUSH} arguments & \np{249076} \\
    Skeletons & \np{248328} \\
    \bottomrule
  \end{tabular}%
\end{table}
The 48.3\,M contract creations involved 2.2\,M different deployment codes, generating a total of 0.5\,M distinct runtime codes.
The removal of meta-data reduces the number of distinct codes by \np[\%]{29}, and the removal of \OP{PUSH} constants by another \np[\%]{22}.

For each family of codes, i.e., for each collection of codes sharing the same skeleton, we pick a single representative and omit the others.
For practical purposes, we prefer deployments where Etherscan lists the corresponding source code.
The longest-lived family consists of two codes implementing an ERC20 token, deployed \np{17333} times over a range of almost 12 million blocks, whereas the most prolific family consists of 20 codes deployed over 12 million times.%
\footnote{%
  The codes of this family, 21 bytes in length, belong to gas token systems.
  When called from the address mentioned in the code, the contracts self-destruct, leading in earlier versions of Ethereum to a gas refund.}
The size of the families seems to follow a Pareto principle: 84\,\% of the families are singletons (the skeleton is uniquely associated with a single runtime code), 15\,\% of the families consist of 2 to 10 codes, whereas at the other end of the spectrum we find a skeleton shared by \np{16372} codes.%
\footnote{%
  These codes are proxy contracts of 45 bytes that forward any incoming call to a fixed address.
  The \np{16372} codes only differ with respect to this hard-coded address, occurring as the argument of a \OP{PUSH} operation.}

We obtain a dataset of \np{248328} runtime codes with distinct skeletons that represent all deployments up to January 13, 2022.
\np[\%]{99.0} of these codes originate from the Solidity compiler (as determined by characteristic byte sequences), with the source code for 46.5\,\% actually available on Etherscan.
\emph{For our temporal analyses, we associate each code with the block number where the first member of the family was deployed.}

Not all bytecodes are proper contracts.
In particular in the early days of the main chain, during an attack, a number of large `contracts' were deployed that served as data repositories for other contracts. 
For some tools, this leads to a noticeable spike in the error rate around block 2.3\,M.

\subsection{Forks Introducing New Operations}\label{sec:new_ops}

\begin{table}
  \centering
  \caption{Forks on Ethereum's main chain introducing new operations}
  \label{tab:forks}
  \makebox[0pt]{%
  \begin{tabular}{lrl}
    \toprule
    name of fork&activated at&new operations\\
    \midrule
    Homestead & 1.150\,M & \OP{DELEGATECALL}\\
    Byzantium & 4.370\,M & \OP{RETURNDATASIZE}, \OP{RETURNDATACOPY},
                           \OP{REVERT}, \OP{STATICCALL}\\
    Constantinople & 7.280\,M & \OP{CREATE2}, \OP{EXTCODEHASH}, \OP{SAR},
                                \OP{SHL}, \OP{SHR}\\
    Istanbul & 9.200\,M & \OP{CHAINID}, \OP{SELFBALANCE}\\
    London & 12.965\,M & \OP{BASEFEE}\\
    \bottomrule
  \end{tabular}}
\end{table}
Over time, Ethereum has seen several updates, so-called forks.
Some of them add new operations to the EVM; \autoref{tab:forks} gives an overview of those relevant to our study.
When interpreting the results of an analysis tool, we have to relate the age of the tool to these forks in order to understand the effect of new operations.

\begin{figure}
  \centering
  \includegraphics[width=.9\columnwidth]{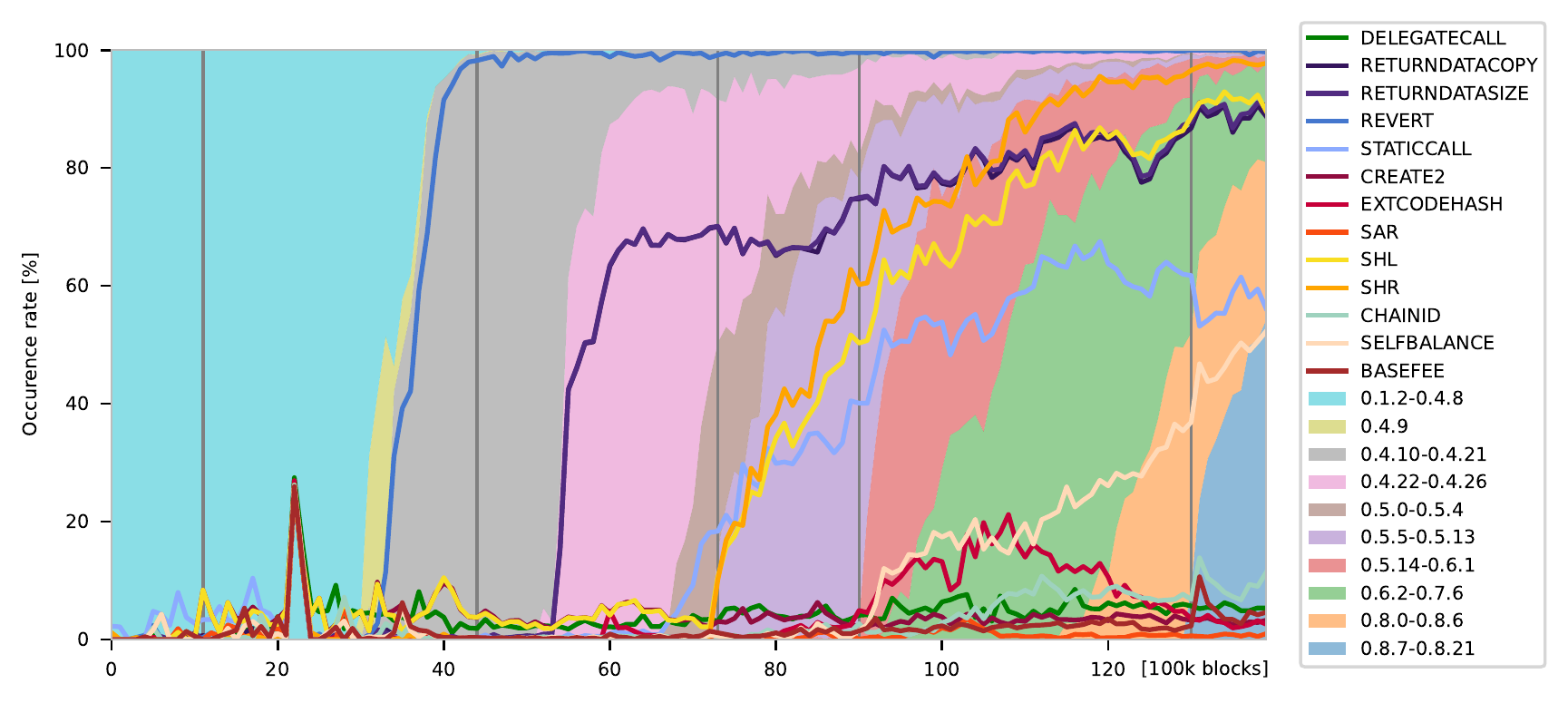}
  \caption{%
    The use of new EVM operations as well as the adoption of new compiler versions over time.
    In the foreground, each line shows the percentage of bytecodes containing a particular operation, per bin of 100\,k blocks.
    The background shows the distribution of compiler versions for each bin.
    \OP{DELEGATECALL} appears from compiler version 0.4.9 onwards, \OP{REVERT} from 0.4.10, \OP{RETURNDATACOPY} and \OP{RETURNDATASIZE} from 0.4.22, \OP{STATICCALL} from 0.5.0, \OP{SAR}, \OP{SHL} and \OP{SHR} from 0.5.5, \OP{SELFBALANCE} from 0.5.14, \OP{CREATE2} from 0.6.2, \OP{CHAINID} from 0.8.0, and \OP{BASEFEE} from 0.8.7 onwards.
    \OP{EXTCODEHASH} is only available via assembly code.}
  \label{fig:ops_evol}
\end{figure}
The actual use of new operations varies.
It depends on the integration of the operations into the Solidity compiler, the adoption of new compiler versions by the contract developers, and on the relevance of the operation.
\autoref{fig:ops_evol} shows the use of compiler versions as well as the share of contracts using a particular operation.
In the remainder of the section, we first explain the origin of the data and then discuss the details of the figure.

For bytecodes with a source code on Etherscan.io (\np[\%]{46.5}), we take the compiler version from there.
For the others, we try to extract it from the embedded meta-data, succeeding for another 22.8\,\% of the bytecodes.
Given that \np[\%]{99} of the bytecodes have been generated by the Solidity compiler, we extend the compiler distribution in each bin of 100\,k blocks from the codes with a known version to the remaining ones.
For the stackplot in the background of \autoref{fig:ops_evol}, this amounts to considering the share of \np[\%]{69.3} bytecodes with version information as 100\,\% of the data.

We group the 100+ compiler versions into 10 ranges, such that from each range to the next, at least one additional operation from~\autoref{tab:forks} is supported, either by Solidity offering a new language element related to the operation, or by the compiler using the new operation internally to generate better bytecode.

It is inherently difficult to determine the parts of a bytecode that are reachable and may get executed.
To obtain a list of the operations in a bytecode, we identify the first code block and extract its operations linearly.
This yields an overapproximation, as data embedded in the code may be mistaken as operations.

In \autoref{fig:ops_evol}, the effect of this overapproximation is visible up to block 3\,M.
Except for \OP{DELEGATECALL}, none of the new operations has been added to the EVM yet, but the plots show their occurrence at a rate of up to \np[\%]{8} nonetheless.
The spike at block 2.3\,M, with all operations seemingly occurring simultaneously at a rate of \np[\%]{28}, is an artefact caused by a large number of `contracts' consisting entirely of data (the by-product of an attack).
The first real phenomenon to observe is the \OP{REVERT} operation (medium blue line), introduced by compiler version 0.4.10 (gray area).
The compiler starts to use \OP{REVERT} for exception handling shortly after 3\,M, even though the operation is added to the EVM only at fork 4.37\,M.
Before the fork, \OP{REVERT} causes an EVM exception just like the invalid opcode used for this purpose until then, but performs a more refined error handling after the fork.

The other operations of fork 4.37\,M are adopted by the Solidity compiler with some delay.
From version 0.4.22 onwards, the pair of \OP{RETURNDATA} operations (lines in two shades of violet, with the plots almost coinciding) is used for most calls and eventually occurs at a rate beyond \np[\%]{80}.
\OP{STATICCALL} (light blue line) gets in use even later, rising to a rate of \np[\%]{60} towards the end of the timeline.

The shift operations \OP{SHL} (yellow line) and \OP{SHR} (orange line) are of particular importance.
At the end of the study period, they occur in almost every contract.
As the function dispatcher at the start of a contract now uses \OP{SHR}, this operation gets executed with virtually every invocation of a contract.
Of the other operations, only \OP{SELFBALANCE} and \OP{EXTCODEHASH} are used to some extent, while \OP{DELEGETECALL}, \OP{CREATE2}, \OP{SAR}, \OP{CHAINID} and \OP{BASEFEE} play a lesser role.%
\footnote{%
  The low number of contracts containing \OP{DELEGATECALL} may seem surprising, given that this operation is essential for hundreds of thousands of proxy contracts.
  But as these contracts show little variety (few skeletons), they are represented by only a small number of contracts in our dataset.}

\newpage
\section{Analysis Tools}
\label{sec:tools}
In this section, we specify the selection of tools and describe those that we use in our analysis.

\subsection{Selection of Tools}

The setting of our study imposes several restrictions on the tools that we can use.
Starting from the \np{140} tools identified by \cite{Rameder2022}, we apply the following selection criteria.
The numbers in parentheses indicate how many tools in the initial collection fulfill the criterion. 
\begin{enumerate}
\item Availability: The tool needs to be publicly available with its source open (83).
\item Input: The tool is able to analyze contracts based on their runtime bytecode alone (73).
  This excludes tools that need access to the application binary interface\footnote{\url{https://docs.soliditylang.org/en/latest/abi-spec.html}}, to the source code, or to a particular state of the blockchain.
\item Findings: The tool offers an automated mode (41) to report weaknesses (79).
  Some systems are tool boxes for the manual analysis of single contracts.
  Moreover, not all tools target weaknesses, but contracts like honeypots or Ponzi schemes. 
\item Interface: The tool can be controlled via a command-line interface.
\item Documentation: There is sufficient documentation to operate the tool.
\end{enumerate}
Criteria without a number were only checked after the ones with a number had been applied.
The selection processes yielded the 12 tools in \autoref{tab:tools}.
\begin{table}
\caption{Tools Selected for the Study. The tools kept in boldface are the ones newly added to the execution framework.}
\label{tab:tools}
\centering
\newcommand\myurl[1]{\href{https://#1}{#1}}
\makebox[0pt]{%
\begin{tabular}{@{}ll@{~~}l@{~~}r@{}}
\toprule
Tool & Reference & URL & Version\\
\midrule
Conkas			&&\myurl{github.com/nveloso/conkas} & \href{https://github.com/nveloso/conkas/commit/6aee0981663dd80310658723c6f48af6ec90cdb6}{6aee09} \\
\textbf{Ethainter}& \cite{Brent2020ethainter}		&\myurl{zenodo.org/record/3760403} & \href{https://hub.docker.com/layers/smartbugs/ethainter/latest/images/sha256-2b26bfb83263c9c4ecf7f62e7dba95709bdfe44f3adaeca55f154c717b2ecaa1?context=explore}{2b26bf} \\ 	
\textbf{eThor}& \cite{Schneidewind2020ethor}	&\myurl{secpriv.wien/ethor/} & \href{https://hub.docker.com/layers/smartbugs/ethor/rmvi20q/images/sha256-6405e0edbf7090a2ce3bca470414766e41b6c3324f1e979f10f0f81622840066?context=explore}{6405e0}\\
\textbf{MadMax}& \cite{grech2018madmax}		&\myurl{github.com/nevillegrech/MadMax} & \href{https://github.com/nevillegrech/MadMax/commit/6e9a6e99c660f0e372e61cf9c7b30be481e047ec}{6e9a6e9} \\
Maian& \cite{nikolic2018finding}					&\myurl{github.com/ivicanikolicsg/MAIAN} & \href{https://github.com/ivicanikolicsg/MAIAN/commit/4bab09ae03f7e27117d653ef8398ec0b2a95f726}{4bab09} \\	
Mythril& \cite{mueller2018smashing}				&\myurl{github.com/ConsenSys/mythril} & \href{https://github.com/ConsenSys/mythril/tree/v0.22.32}{0.22.32} \\ 	
Osiris& \cite{FerreiraTorres2018osiris}				&\myurl{github.com/christoftorres/Osiris} & \href{https://github.com/christoftorres/Osiris/commit/ff382832e5cdc3d6ca95c5e35197da4e358a8ed4}{ff3828}\\	
Oyente& \cite{Luu2016}						&\myurl{github.com/enzymefinance/oyente} & \href{https://github.com/enzymefinance/oyente/commit/480e725aa02f7ee92af9227216e57d2193e0f999}{480e725} \\ 	
\textbf{Pakala}	&&\myurl{github.com/palkeo/pakala} & \href{https://pypi.org/project/pakala/1.1.10/}{1.1.10} \\	
Securify& \cite{Tsankov2018securify}				&\myurl{github.com/eth-sri/securify} & \href{https://hub.docker.com/layers/smartbugs/security/usolc/images/sha256-d367b17b6f1ad898a16cf5d663bc95eaf2cefa5de8779590d31575493f9de799?context=explore}{d367b1}\\
\textbf{teEther}& \cite{Krupp2018teether}		&\myurl{github.com/nescio007/teether} & \href{https://github.com/nescio007/teether/commit/04adf568b5dcaf9dba0c655e3f234d19764cb603}{04adf5} \\
\textbf{Vandal}& \cite{brent2018vandal}		&\myurl{github.com/usyd-blockchain/vandal} & \href{https://github.com/usyd-blockchain/vandal/commit/d2b004326fee33920c313e64d0970410b1933990}{d2b004}\\
\bottomrule
\end{tabular}}
\end{table}

\subsection{Synopsis of Tools}

\ITEM{Conkas} uses the third-party component Rattle to construct a control flow graph and to lift the bytecode to static single assignments.
Then it executes this intermediate representation symbolically and checks the execution traces for patterns that indicate weaknesses.

\ITEM{Ethainter} uses taint analysis to detect whether attacker-injected data reaches critical operations.
It relies on the component Gigahorse that abstracts the bytecode to Datalog rules.
The Datalog program, augmented by rules for the addressed weaknesses, is translated to a \Cpp\ program (using the tool Soufflé), which is compiled to machine code that performs the actual analysis.

\ITEM{eThor} attempts to either prove that the contract is not reentrant and thus is not susceptible to a reentrancy attack, or to find a trace with a reentrant call that indicates that the contract may be vulnerable.
eThor lifts the bytecode to constrained Horn clauses that constitute a sound abstraction of the EVM semantics.
The Horn clauses are then translated into the constraint language of the SMT solver Z3 that does the final analysis.
Various optimizations like the unfolding of Horn clauses are crucial to make the analysis feasible.

\ITEM{MadMax} concentrates on gas-related vulnerabilities.\footnote{%
  EVM instructions consume gas proportional to the time and storage they need.
  Each transaction is endowed with a limit on the total gas it may use, ensuring that each contract terminates.}
Like Ethainter, it uses Gigahorse and Soufflé to lift the bytecode to Datalog.

\ITEM{Maian} executes the EVM bytecode symbolically, relying on the SMT solver Z3 to check the satisfiability of path conditions.
When analyzing source or deployment code, Maian additionally validates the detected weaknesses by deploying the contract on a private blockchain and attacking it with the transactions computed by the SMT solver.
For runtime code this validation step has to be omitted, as there is not enough information for deployment.

\ITEM{Mythril} uses symbolic execution, SMT solving (Z3) and taint analysis to detect a variety of security vulnerabilities.
To increase code coverage, it applies concolic execution that alternates between symbolic execution and runs with concrete values.
It is the only tool in the selection that is actively maintained.

\ITEM{Oyente} constructs a control-flow graph and then executes the contract symbolically.
The execution traces are checked for patterns characteristic of certain weaknesses.
In a final validation step, certain false positives are eliminated before reporting the findings.
Symbolic path constraints and the SMT solver Z3 are used to prune the search space.

\ITEM{Osiris} extends Oyente by adding modules for further weaknesses.

\ITEM{Pakala} executes the bytecode symbolically, without constructing a control flow graph first.
It proceeds in two phases.
First, it collects transactions that lead to state changes or Ether transfers.
Then, the transactions are combined in varying sequences to find one that extracts more Ether than was invested with the transaction.

\ITEM{Securify} constructs a control flow graph and lifts the bytecode to single static assignments.
It uses Souffl\'e to derive semantic facts from inference rules specified in Datalog that describe the data and control flow dependencies.
These facts are then checked against a set of compliance and violation patterns written in a logic-based domain-specific language.

\ITEM{teEther} constructs a control flow graph and searches for paths that lead to critical instructions, with arguments controllable by an attacker.
Symbolic execution translates these paths into constraints for the SMT solver Z3.
The solutions computed by Z3 can then be turned into exploits, i.e., into transactions that trigger the suspected vulnerabilities.

\ITEM{Vandal} constructs a control flow graph and lifts the bytecode to single static assignments.
This intermediate representation is translated to logic relations that represent the semantics of the initial bytecode.
Weaknesses are specified as Datalog rules.
Soufflé synthesizes executable programs that read the logic relations and perform the security analysis.

\subsection{Maintenance Aspects}

\begin{table}
\caption{Maintenance Aspects of Tools (checked in February 2023)}
\setlength{\tabcolsep}{3pt}
\label{tab:maintenance}
\centering
\begin{tabular}{@{}lr
  |rr|rrc|l@{}}
  \toprule
  \multirow{2}{*}{Tool} & \multirow{2}{*}{Year}
  & \multicolumn{2}{c|}{ Github issues} & \multicolumn{3}{c|}{Commits}   & \multirow{2}{*}{Based on}    \\
                        &  & 
                                 open & closed &\faUser&number&   last  & \\
  \midrule
  Conkas      & 2021 
  &  8 &                    2 &  1 &   3 & 05.2022 & Rattle \\
  Ethainter   & 2020 
  &    &                      &    &     &         & Gigahorse, Souffl\'e \\
  eThor       & 2020 
  &    &                      &    &     &         & HoRSt, Souffl\'e \\
  MadMax      & 2020 
  &  2 &                    5 &  7 & 910 & 06.2021 & Gigahorse, Souffl\'e \\
  Maian       & 2018 
  & 27 &                    8 &  1 &  15 & 03.2018 & \\
  Mythril     & 2020 
  & 79 &          709 & 74 & \np{4785} & 02.2023 & \\
  Osiris      & 2018 
  &  4 &                    2 &  1 &   5 & 09.2018 & Oyente \\
  Oyente      & 2017 
  & 63 &                  126 & 23 & 848 & 11.2020 & \\
  Pakala      & 2018 
  &  0 &                   12 &  1 & 197 & 03.2020 & \\
  Securify    & 2019 
  & 18 &                   53 &  6 & 155 & 09.2019 & Souffl\'e \\
  teEther     & 2018 
  &  2 &                   18 &  1 &  21 & 07.2021 & \\
  Vandal      & 2018 
  & 27 &                   10 & 10 & 870 & 07.2020 & Souffl\'e \\
\bottomrule
\end{tabular}
\end{table}

\autoref{tab:maintenance} lists statistics related to the effort put into keeping the tools up to date.
Some tools show a small number of commits and closed issues only, and thus seem unmaintained.
But even tools with several hundred commits became unmaintained at some point.
Judged by the last code commits, Mythril is the only tool actively maintained: it has 74 contributors, \np{4785} commits, and 709 issues resolved.
For our study, maintenance mainly boils down to the question which EVM operations the tools are actually able to handle.

\subsection{Supported EVM Operations}\label{sec:supported_ops}

\begin{table}
  \let\Y\YES
  \newcommand\n{}%
  \caption{EVM operations supported by the tools. A question mark indicates that
  the operation seems to be handled by the code, but causes an exception.}
  \label{tab:operations}
  \centering
  \makebox[0pt]{%
  \begin{tabular}{lc|cccc|ccccc|cc|c}
    since block
    & 1.15\,M
    & \multicolumn{4}{c|}{4.37\,M}
    & \multicolumn{5}{c|}{7.28\,M}
    & \multicolumn{2}{c|}{9.069\,M}
    & 12.965\,M \\
    & \SW{\OP{DELEGATECALL}}
    & \SW{\OP{RETURNDATACOPY}} 
    & \SW{\OP{RETURNDATASIZE}}
    & \SW{\OP{REVERT}}
    & \SW{\OP{STATICCALL}}
    & \SW{\OP{CREATE2}}
    & \SW{\OP{EXTCODEHASH}}
    & \SW{\OP{SAR}}
    & \SW{\OP{SHL}}
    & \SW{\OP{SHR}}
    & \SW{\OP{CHAINID}}
    & \SW{\OP{SELFBALANCE}}
    & \SW{\OP{BASEFEE}}\\
    \midrule
    Conkas   &\Y&\Y&? &\Y&\Y&? &? &? &? &? &? &? &\n\\\rowcolor{lightgray}
    Ethainter&\Y&\Y&\Y&\Y&\Y&\Y&\n&\Y&\Y&\Y&\n&\n&\n\\
    eThor    &\n&\Y&\Y&\Y&? &? &? &? &\Y&? &\n&\n&\n\\\rowcolor{lightgray}
    MadMax   &\Y&\Y&\Y&\Y&\Y&\Y&\Y&\Y&\Y&\Y&\Y&\Y&\n\\
    Maian    &\Y&\n&\n&\Y&\n&\n&\n&\n&\n&\n&\n&\n&\n\\\rowcolor{lightgray}
    Mythril  &\Y&\Y&\Y&\Y&\Y&\Y&\Y&\Y&\Y&\Y&\Y&\Y&\Y\\
    Osiris   &\Y&\Y&\Y&\Y&\n&\n&\n&\n&\n&\n&\n&\n&\n\\\rowcolor{lightgray}
    Oyente   &\Y&\Y&\Y&\Y&\Y&\n&\n&\n&\n&\n&\n&\n&\n\\
    Pakala   &\Y&\Y&\Y&\Y&\n&\n&\n&\Y&\Y&\Y&\Y&\Y&\n\\\rowcolor{lightgray}
    Securify &\Y&\Y&\Y&\Y&\Y&\Y&\Y&\Y&\Y&\Y&\n&\n&\n\\
    teEther  &\Y&\Y&\Y&\Y&\Y&\n&\n&\Y&\Y&\Y&\n&\n&\n\\\rowcolor{lightgray}
    Vandal   &\Y&\Y&\Y&\Y&\Y&\Y&\Y&\Y&\Y&\Y&\n&\n&\n\\
    \bottomrule
  \end{tabular}}
\end{table}

We say that a tool supports an EVM operation if it models at least the effect of the operation on the stack, by removing and adding an appropriate number of elements.
This way the analysis of the current execution path can proceed, even if little may be known about the state after the operation.
EVM operations may be unsupported either because of having been omitted deliberately or because of having been introduced by a fork after the tool was released (see \autoref{sec:new_ops}).
Tools handle unsupported operations by either stopping the analysis of the current path with an error message, by reverting the computation like the EVM would do for an unused opcode, or by aborting with an exception.

The effect of unsupported operations depends on the type of property checked for.
Most properties are existential: A bytecode satisfies the property if \emph{some} execution path satisfies a characteristic condition.
In such cases an unhandled operation will reduce the number of paths that can be checked.
The tool remains sound, but its detection rate diminishes with the number of paths it cannot check.
A few properties are universal in nature: The property holds for a bytcode if a condition is met by \emph{all} execution paths.
Here, any path that cannot be checked is a threat to the validity of the result, making the method unsound.
\emph{Ether lock} is an example for both property types: The weakness is present if there is at least one execution path that increases the balance of the contract (the contract accepts Ether), whereas all execution paths have the property that they do not decrease the balance (the contract offers no withdraw functionality).

To interpret the analysis results later on, we determine the operations supported by each tool.
The data in \autoref{tab:operations} was obtained by inspecting the source code as well as by executing the tools on a collection of crafted contracts where each consists of one of the listed operations, with some \OP{PUSH} operations preparing the stack. 
A checkmark in the table indicates that the tool supports the operation, while a question mark means that the tool supports the operation according to its source code, but fails on the corresponding crafted contract.

The tools usually support the operations up to the most recent fork before their last update.
Most tools analyze single contracts only, therefore they handle interactions with other contracts by not supporting the operations \OP{DELEGATECALL}, \OP{STATICCALL}, \OP{CREATE2} and \OP{EXTCODEHASH} at all, or if they do, by invalidating those parts of the state that might have been affected by the operation.
Conkas and eThor stick out as handling several operations in their source code that lead to exceptions during execution.
Maian is older than it looks: The tool was published some time after the fork at block 4.37\,M, yet it does not support the two \OP{RETURNDATA} operations. 
\newpage
\section{Execution Framework}
\label{sec:execution}
For the large-scale execution of our study, we had the choice between two frameworks: SmartBugs~\citep{ferreira2020smartbugs} and USCV~\citep{Ji2021}, both operating on Solidity level.
We decided on the former, as SmartBugs is better maintained (c.f.\ the last paragraph of \autoref{sec:related}) and contained more of the tools we were interested in.

First, we adapted SmartBugs to accept bytecode as input, and updated the Docker images of the tools accordingly.
Second, we integrated six further tools (kept in boldface in \autoref{tab:tools}).
The most laborious part was the \emph{output parsers}.
For each tool, a dedicated parser scans the output of the tool to identify the result of the analysis, to detect anomalies, and to discard irrelevant messages.
For each run of a tool on a bytecode, the parser reports a list of \emph{findings} (tags identifying the detected properties), a list of \emph{errors} (conditions checked for and reported by the tool), a list of \emph{fails} (low-level exceptions not adequately handled by the tool), and a list of \emph{messages} (any other noteworthy information issued by the tool).
For a detailed description of the enhanced framework see \citep{diangelo2023smartbugs}.

\ITEM{Choice of Parameters.}
\cite{Ren2021} show that the choice of parameters strongly affects the results, especially when the timeout is below 30 minutes per contract.
We set the maximal runtime to \np[s]{1800} wall time, with 1.5 CPUs assigned to each run.
If a tool offers a timeout parameter, we communicate the runtime minus a grace period to allow the tool to terminate properly.
Conkas, eThor, Maian, Securify, teEther and Vandal offer no such parameter and are stopped by the external timer.

As there is a tradeoff between the memory limit per process and the number of processes run in parallel, we aimed at providing sufficient but not excessive memory.
Based on an initial test with \np{500} randomly selected contracts, we set the memory limit to \np[GB]{20} for eThor, Pakala, Securify and teEther, and to \np[GB]{4} for all other tools.
We reran tasks with a limit of \np[GB]{32} if they had failed with a segmentation fault or a memory problem.

\ITEM{Machine.}
We used a server with an AMD EPYC7742 64-Core CPU and \np[GB]{512} of RAM.
\autoref{tab:resources} gives an overview of the computation time, memory usage, and memory fails before and after the rerun with \np[GB]{32}.

\begin{table}
  \centering
  \caption{Resource Consumption and Out-of-memory (OOM) conditions}
  \label{tab:resources}
  \makebox[0pt]{%
  \begin{tabular}{llrrr|rrl}
    \toprule
    \multirow{2}{*}{Tool} & \multirow{2}{*}{Language} & \multirow{2}{*}{Runtime} & \multirow{2}{*}{Memory}
    & \multirow{2}{*}{OOM} & \multicolumn{2}{c}{Re-runs with \np[GB]{32}} \\
    & & & & & Runtime & OOM \\
 \midrule   
 Conkas      & Python &  \np[d]{338} &  \np[GB]{4} &       319  &   \np[d]{5} &        44  \\
 Ethainter   & Python, Datalog &  \np[d]{205} &  \np[GB]{4} &         0  &             &            \\
 eThor       & Java & \np[d]{1651} & \np[GB]{20} &       534  &   \np[d]{7} &       369  \\
 MadMax      & Python, Datalog &   \np[d]{61} &  \np[GB]{4} &         0  &             &            \\
 Maian       & Python &  \np[d]{138} &  \np[GB]{4} &        91  &   \np[d]{1} &        28  \\
 Mythril     & Python & \np[d]{1926} &  \np[GB]{4} &        50  &   \np[d]{1} &        18  \\ 
 Osiris      & Python &  \np[d]{475} &  \np[GB]{4} &       371  &   \np[d]{6} &        14  \\
 Oyente      & Python &   \np[d]{96} &  \np[GB]{4} &  \np{2229} &   \np[d]{5} &  \np{1929} \\
 Pakala      & Python & \np[d]{3204} & \np[GB]{20} &       421  &   \np[d]{6} &       333  \\
 Securify    & Java, Datalog &  \np[d]{460} & \np[GB]{20} &         0  &             &            \\
 teEther     & Python & \np[d]{1678} & \np[GB]{20} & \np{47794} & \np[d]{559} & \np{40306} \\
 Vandal      & Python, Datalog &  \np[d]{172} &  \np[GB]{4} &  \np{2051} &  \np[d]{23} &  \np{1142} \\
  \bottomrule
  \end{tabular}}
\end{table}

\newpage
\section{Weaknesses}
\label{sec:weaknesses}
In this section, we describe the weaknesses considered in our study as well as the taxonomy used and the mapping of the tool findings to the taxonomy.

\subsection{Vulnerability Detection vs.\ Weakness Warning}
According to the Common Weakness Enumeration, \href{https://cwe.mitre.org/about/index.html}{cwe.mitre.org}, weaknesses are \emph{flaws, faults, bugs, or other errors in software or hardware implementation, code, design, or architecture that if left unaddressed could result in systems, networks, or hardware being vulnerable to attack.}

The tools in our study report findings with varying degrees of certainty, from warnings about potential weaknesses to exploits demonstrating the existence of a vulnerability or, more rarely, proofs guaranteeing their absence.
As proving the absence or presence of software properties is difficult, most tools employ heuristics, usually favoring a higher number of false positives over the possibility to overlook an actual vulnerability.
Such tools issue warnings and leave the final assessment to the user.

\subsection{Synopsis of Weaknesses}
\label{sec:synopsis_weaknesses}
\ITEM{Integer Overflow and Underflow (SWC 101)}
Integer over- and underflow weaknesses arise in situations, where the result of an arithmetic operation exceeds the admissible range and the system performs a silent wrap-around instead of throwing an exception.
The wrong result may lead to an unexpected behavior or a security breach.

\ITEM{Unchecked Call Return Value (SWC 104)}
This weakness arises when smart contracts do not properly validate the return value of an external contract call, such that an unusual behavior of the call goes unnoticed.

\ITEM{Unprotected Ether Withdrawal (SWC 105)}
Unprotected Ether withdrawal weaknesses occur when smart contracts allow unauthorized parties to withdraw Ether without proper access controls, risking financial loss.

\ITEM{Unprotected \OP{SELFDESTRUCT} Instruction (SWC 106)}
This weakness arises when a \OP{SELFDESTRUCT} instruction is not properly guarded such that an attacker can trigger the destruction of the contract, potentially resulting in the loss of funds or a disruption of services.

\ITEM{Reentrancy (SWC 107)}
A reentrancy weakness occurs when a contract calls another one without updating its internal state beforehand.
If the callee calls the caller back, the latter may be in an inconsistent state.

\ITEM{Assert Violation (SWC 110)}
Assertions are sanity checks that are meant to hold for every run.
An assert violation means that such a check can be made to fail for certain inputs, which can lead to unexpected termination and loss of funds.

\ITEM{Delegatecall to Untrusted Callee (SWC 112)}
Code invoked by a \OP{DELEGATECALL} instruction operates on the caller's storage and funds.
This weakness means that an attacker is able to manipulate the caller's state by controlling the code that is invoked.

\ITEM{DoS with Failed Call (SWC 113)}
A failing call to an external contract may prevent subsequent actions from taking place.
A buggy or malicious callee may cause a DoS with the caller.

\ITEM{Transaction Order Dependence (SWC 114)}
This weakness arises when a contract's behavior depends on the order in which transactions are mined, leading to inconsistencies and potential security issues.

\ITEM{Authorization through tx.origin (SWC 115)}
This weakness means that a contract relies on tx.origin for authorization, which can be exploited by a man-in-the-middle attack to bypass access controls.

\ITEM{Block Values as a Proxy for Time (SWC 116)}
This weakness results from using block-related values as a substitute for precise timing, as these values can be manipulated by attackers.

\ITEM{Weak Sources of Randomness from Chain Attributes (SWC 120)}
Seeding random number generators with chain attributes leads to weak randomness, as these attributes can be predicted or even manipulated by attackers.

\ITEM{Write to Arbitrary Storage Location (SWC 124)}
This weakness occurs when an attacker is able to write to an unintended storage location, e.g.\ by overflowing one data structure into the next in storage.

\ITEM{Arbitrary Jump with Function Type Variable (SWC 127)}
Solidity supports function types.
By low level manipulations of variables that hold a function, control can be handed over to code other than the function, leading to unintended execution paths.

\ITEM{DoS With Block Gas Limit (SWC 128)}
The resource consumption of every call to a contract is limited by the gas supplied.
But this gas limit is capped itself by the block gas limit.
Once the call to a contract requires more gas than that, e.g.\ because of looping over a data structure that has been grown too big by an attacker, the contract may become inoperable.

\ITEM{Ether Lock}
This weakness means that a contract accepts Ether without offering the functionality to withdraw it, thus locking any funds sent to it.
    
\ITEM{Callstack Depth Bug}
Originally, the number of nested calls was limited to 1024.
Calling a contract at this limit would make any further nested calls fail unexpectedly, leading to a potentially harmful situation for the called contract.
As early as block 2.463\,M, this limit was replaced by a better mechanism that made the weakness related to the callstack depth limit obsolete.

\subsection{Mapping of Tool Findings}\label{sec:mapping}

\ITEM{Taxonomy.}
To compare the tools regarding their ability to detect weaknesses, we need a taxonomy with an adequate granularity.
Since there is no established taxonomy of weaknesses for smart contracts, previous studies~\citep{Chen2020survey, Tang2021, Wang2021survey, Kushwaha2022vuls, Rameder2022, tolmach2022survey, Zhou2022state} not only summarize potential issues, but also structure them with respect to their own taxonomies, none of which is compelling or widely used.

Among the community projects, there are two popular taxonomies: the DASP (Decentralized Application Security Project) TOP 10~\footnote{https://dasp.co} from 2018, which features 10 categories, and the SWC registry (Smart Contract Weakness Classification and Test Cases)~\footnote{https://swcregistry.io} with 37 classes, last updated 2020.
As for DASP, two categories, \textit{Access Control~(2)} and \textit{Other~(10)}, are quite broad, while \textit{Short Address~(9)} is checked by hardly any tool.
Moreover, \textit{DOS~(5)} and \textit{Bad Randomness~(6)} are effects that may be the result of various causes, and most tools detect causes rather than consequences.

The SWC registry is more granular as it offers several classes for the broad categories \textit{Access Control} and \textit{DOS}.
Moreover, most of its categories match relevant findings of the tools.
Therefore, we select this taxonomy as the basis of our comparison.

\begin{table}
\caption{Weakness Classes and Reports}
\label{tab:vuls}
\centering
\makebox[0pt]{%
\begin{tabular}{@{}lrrr@{}}
\toprule
Weaknesses& SWC & Frequency   & Tools \\
\midrule
\SWC{101} & 101 & \np{123255} & 4 \\
\SWC{104} & 104 & \np{183277} & 4 \\
\SWC{105} & 105 &  \np{17916} & 7 \\
\SWC{106} & 106 &   \np{7310} & 4 \\
\SWC{107} & 107 & \np{184610} & 7 \\
\SWC{110} & 110 &  \np{62246} & 1 \\
\SWC{112} & 112 &   \np{1352} & 3 \\
\SWC{113} & 113 &  \np{20202} & 2 \\
\SWC{114} & 114 &  \np{35875} & 4 \\
\SWC{115} & 115 &   \np{8669} & 2 \\
\SWC{116} & 116 &  \np{46372} & 4 \\
\SWC{120} & 120 &   \np{8733} & 1 \\
\SWC{124} & 124 &   \np{8826} & 2 \\
\SWC{127} & 127 &   \np{2474} & 1 \\
\SWC{128} & 128 &   \np{8863} & 1 \\
\bottomrule
\end{tabular}}
\end{table}

\ITEM{Findings mapped.}
The tools report \np{82} different findings, of which we can map \np{56} to one of the 37 classes of the SWC taxonomy (see  \autoref{tab:mapping} in the appendix).
In total, the tools cover \np{15} SWC classes.
\autoref{tab:vuls} lists the weakness classes, the accumulated number of findings that the tools report, and the number of tools that address the weakness.
\autoref{tab:tool_vuls} gives an overview of the coverage of the SWC classes by tool.

\begin{table}
\caption{SWC Classes Detected by Tools}
\setlength{\tabcolsep}{2pt}
\label{tab:tool_vuls}
\centering
\makebox[0pt]{%
\begin{tabular}{@{}lcccccccccccccccc}
 \toprule
 Tool		& 101	& 104	& 105	& 106	& 107	& 110	& 112	& 113	& 114	& 115	& 116	& 120	& 124	& 127	& 128	& total
\\ \midrule
Conkas		& \YES	& \YES	&		&		& \YES	&		&		&		& \YES	&		& \YES	&		&		&		& 		& 5 \\
\rowcolor{lightgray}
Ethainter	&		&		& \YES	& \YES	&		&		& \YES	&		&		&		&		&		& \YES	&		&		& 4 \\
eThor 		&		&		&		&		& \YES	&		&		&		&		&		&		&		&		&		& 		& 1 \\
\rowcolor{lightgray}
MadMax		& \YES	&		&		&		&		&		&		& \YES	&		&		&		&		&		&		& \YES	& 3 \\
Maian		&		&		& \YES	& \YES	&		&		&		&		&		&		&		&		&		&		& 		& 2 \\
\rowcolor{lightgray}
Mythril 		& \YES	& \YES	& \YES	& \YES	& \YES	& \YES	& \YES	& \YES	&		& \YES	& \YES	& \YES	& \YES	& \YES	& 		& 13 \\
Osiris 		& \YES	&		&		&		& \YES	&		&		&		& \YES	&		& \YES	&		&		&		& 		& 4 \\
\rowcolor{lightgray}
Oyente 		&		&		&		&		& \YES	&		&		&		& \YES	&		& \YES	&		&		&		& 		& 3 \\
Pakala 		&		&		& \YES	&		&		&		& \YES	&		&		&		&		&		&		&		& 		& 2 \\
\rowcolor{lightgray}
Securify 		&		& \YES	& \YES	&		& \YES	&		&		&		& \YES	&		&		&		&		&		& 		& 4 \\
teEther 		&		&		& \YES	&		&		&		&		&		&		&		&		&		&		&		& 		& 1 \\
\rowcolor{lightgray}
Vandal 		&		& \YES	& \YES	& \YES	& \YES	&		&		&		&		& \YES	&		&		&		&		& 		& 5 \\
\bottomrule
\end{tabular}}
\end{table}

When a tool reports a finding, we assume that it is not invalidated by an accompanying error condition, a low coverage of the bytecode, or a timeout.
However, we note errors, timeouts, and unhandled conditions (fails).

\ITEM{Findings omitted.}
In order not to count the same weakness twice, we exclude redundant and intermediate findings.
Moreover, some findings state the absence of a weakness and thus should not be counted as a weakness.
Altogether, we exclude seven findings:
For eThor the positive finding \emph{secure} (from reentrancy);
for Maian the intermediate finding \emph{accepts\_Ether} and the positive findings \emph{no\_Ether\_leak}, \emph{no\_Ether\_lock}, \emph{not\_destructible};
for Osiris the redundant finding \emph{arithmetic\_bug} (as it is doubled by a more specific one); 
and for Vandal the preliminary finding \emph{checked\_call\_state\_update}.

\newpage
\section{RQ1 Abstraction}\label{sec:rq1}

To validate our hypothesis that bytecodes with the same skeleton (i.e., members of the same code family) behave similarly regarding bytecode analysis, we randomly select \np{1000} bytecodes from all runtime bytecodes \emph{not} in our data set.
By construction, these codes belong to families with at least two members.
The selected bytecodes happen to belong to \np{620} families.
We add the corresponding \np{620} representatives from our data set, obtaining a dataset with \np{1620} bytecodes and \np{620} families with \np{2} to \np{64} members per family.

When running analysis tools on different members of the same family, we expect nearly identical results with small variations due to differences in runtimes (e.g.\ one run timing out while the other one finishes just in time with some finding) or due to the effect of different constants when solving constraints.
In particular, we do not expect the meta-data injected by the Solidity compiler to affect the result, as it is interpreted neither as code nor as data during execution.
To confirm this, we also consider a copy of our \np{1620} bytecodes, where we replace all meta-data sections with zeros.

\begin{table}
\caption{Code Families with Diverging Results [\%]}
\label{tab:variability}  
\centering
  \newcommand\hlb{\cellcolor{midblue}}%
  \begin{tabular}{@{}lrrrr@{}}
    \toprule
    \multirow{2}{*}{Tool}    & \multicolumn{2}{c}{with meta-data} & \multicolumn{2}{c}{without meta-data} \\
    & findings & all fields & findings & all fields \\
    \midrule
Conkas&1.5&1.9&1.3&1.9\\
Ethainter&0.0&0.2&0.0&0.3\\
MadMax&0.0&0.0&0.0&0.5\\
Mythril&0.2&0.6&1.0&1.6\\
Pakala&0.0&1.8&0.0&1.8\\
Securify&0.0&0.0&0.2&0.5\\
teEther&0.0&2.3&0.0&2.3\\
    \midrule
Oyente&0.2&\hlb20.0&0.3&0.5\\
Osiris&1.6&\hlb21.3&0.8&1.3\\
    \midrule
eThor&\hlb2.9&\hlb8.2&1.0&1.5\\
Vandal&\hlb3.2&\hlb4.7&0.2&2.1\\
Maian&\hlb11.3&\hlb11.5&0.5&0.6\\
    \bottomrule
  \end{tabular}
\vspace{-1em}
\end{table}

\autoref{tab:variability} shows the result of running all tools on the bytecodes with and without meta-data.
Columns two and four give the percentage of the 620 families for which the findings differ within the family, whereas columns three and five consider all data collected by the output parsers, including errors, fails, and messages.
If we assume that the various effects influencing the output give rise to a normal distribution, then for a confidence level of \np[\%]{95}, the sample size of 620 yields a margin of error of \np[\%]{1.5} for the smaller values in the table and of \np[\%]{3.2} for the larger ones.

The seven tools on top behave essentially as predicted.
For Conkas, the rate of \np[\%]{1.5} corresponds to \np{9} families with divergent findings.
These differences are related to warnings about integer under- and overflows, and may indeed be the result of different constants in the codes of a family.
Observe that for these seven tools, there is hardly any difference between the two datasets, with and without meta-data.

Osiris and Oyente seem remarkable, as we find \np[\%]{20} discrepancies in the output.
Oyente starts its analysis by disassembling the entire bytecode.
It issues the warning `incomplete push instruction' when stumbling upon a supposed \OP{PUSH} instruction near the end of the meta-data that is followed by too few operand bytes.
These spurious messages disappear when removing the meta-data, but otherwise do not affect the analysis.
Osiris reuses Oyente's code and inherits this anomaly.

eThor also scans the entire bytecode.
When encountering an unknown instruction, it issues a warning and ignores the remaining code.
Like with Oyente, these messages mostly disappear when removing the meta-data.
However, unlike Oyente, the meta-data influences the result of the analysis, as can be observed by \np[\%]{2.9} vs.\ \np[\%]{1.0} differences in the findings for code with vs.\ no meta-data.
In each of these cases, the analysis times out for some member(s) of the family but terminates with identical results for the others.
We did not research the cause for these discrepancies but suspect that it may be comparable to the situation of Vandal.

Vandal constructs a control flow graph for the entire bytecode and decompiles it to an intermediate representation.
Sometimes, the tool gets lost during this initial phase and times out.
The situation improves when removing irrelevant parts like the meta-data.
However, as Vandal interprets the addresses of all code sections relative to the beginning of the bytecode, even if they belong to a different contract (see the discussion on the structure of bytecode in \autoref{sec:creation}), we still see differences regarding errors and fails.

Maian starts by scanning the entire bytecode for certain instructions, like \OP{SELFDESTRUCT}.
Not detecting the opcode anywhere lets Maian immediately conclude certain properties, whereas finding the opcode triggers a reachability analysis that may remain inconclusive.
This sensitivity to single bytes yields divergent results for \np{70} families.
For example, Maian may detect non-destructibility for one code and fail to do so for another one in the same family.
Removing the meta-data gets rid of these divergences almost entirely.

\begin{mdframed}[style=mpdframe]
\textbf{Observation 1.}
Treating bytecodes with the same skeleton as equivalent works for 9 out of 12 tools without reservations.
Three tools unexpectedly analyze the meta-data, leading to minor output variations. 
Therefore, skeletons can be regarded as a suitable abstraction for large-scale analyses aimed at the big picture.
Removing the meta-data prior to analysis may improve the performance of some tools (while not harming others).
\end{mdframed}

\newpage
\section{RQ2 Weakness Detection over Time}\label{sec:rq2}
\begin{figure}
\centering
\includegraphics[width=.9\columnwidth]{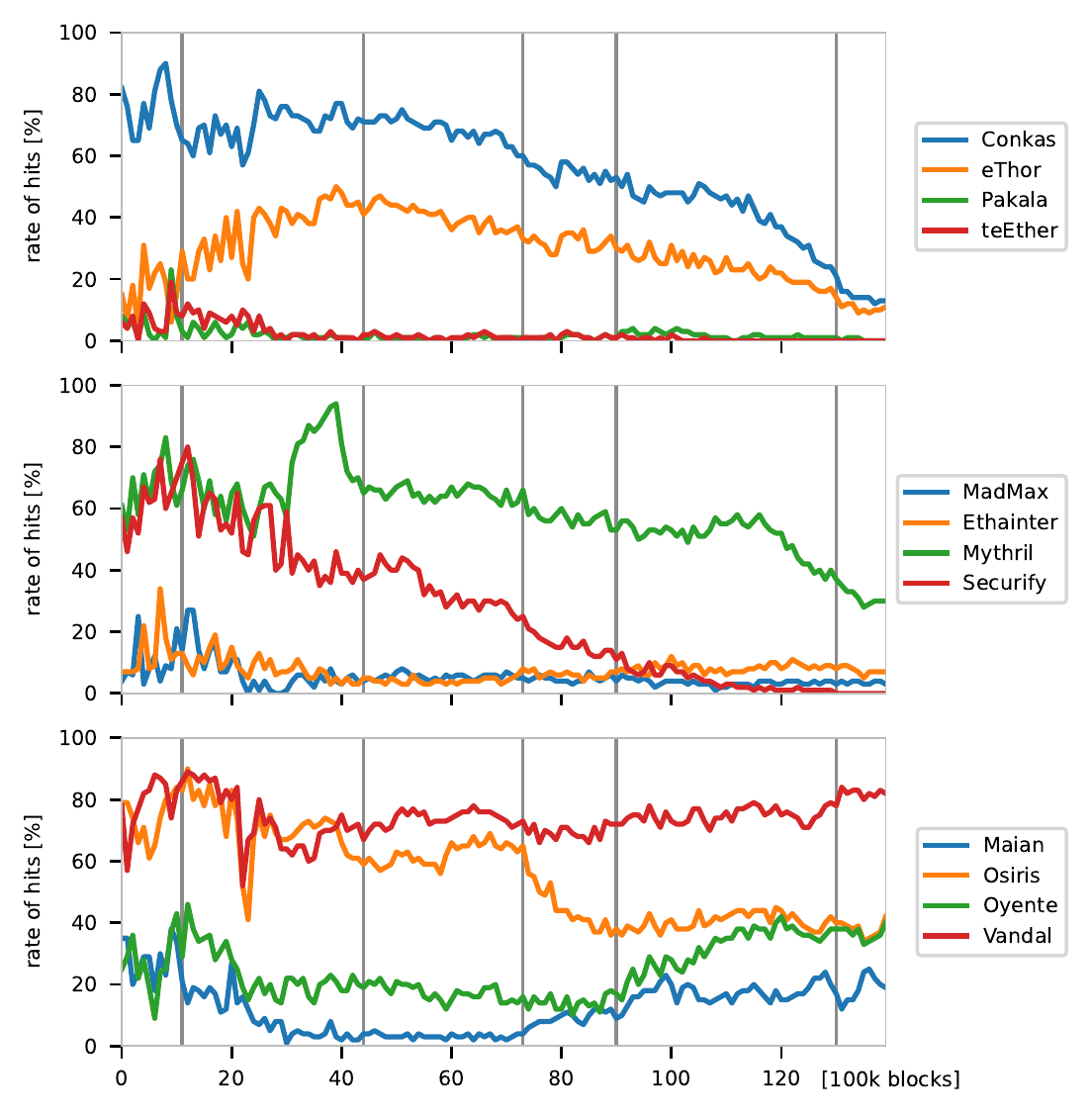}
\caption{Accumulated findings per tool over time. Each data point shows the percentage of bytecodes for which the tool reports a weakness, in bins of 100\,k blocks.}
\label{fig:findings}
\includegraphics[width=.9\columnwidth]{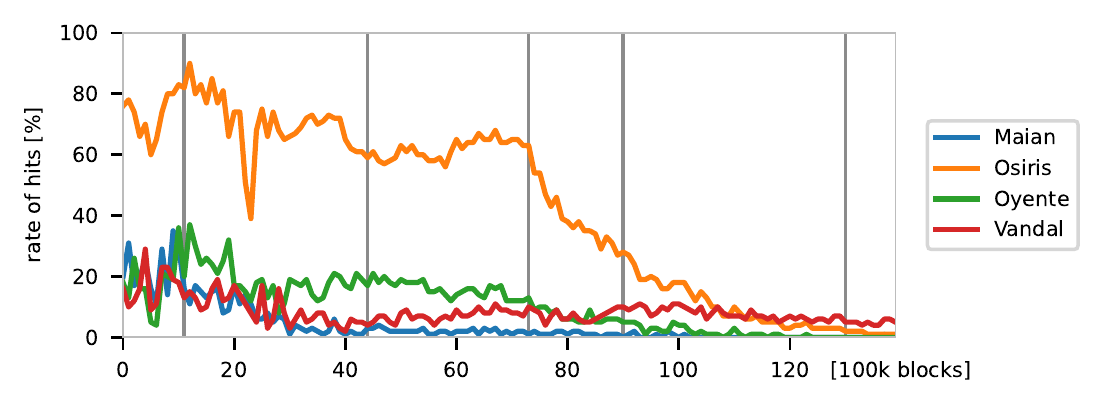}
\caption{Accumulated findings for Maian, Osiris, Oyente and Vandal. Compared to the third plot in \autoref{fig:findings}, some spurious findings have been omitted (see the text for details).}
\label{fig:findings2}
\end{figure}

\begin{figure}
  \centering
  \includegraphics[width=.9\columnwidth]{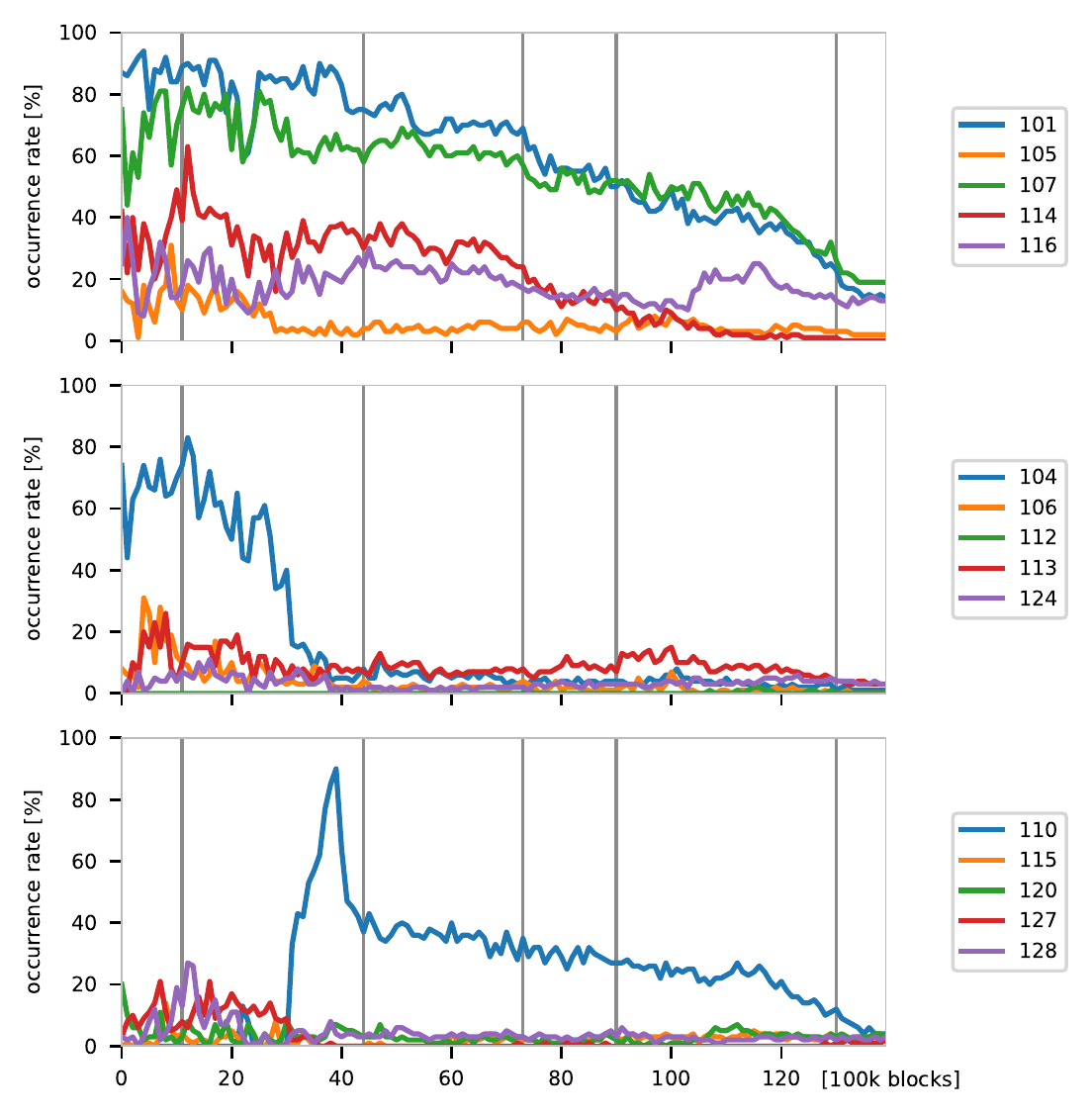}
  \caption{SWC classes over time. Each data point shows the percentage of bytecodes flagged with a specific weakness, in bins of 100\,k blocks.}
  \label{fig:swc_classes}
\end{figure}

In this section, we portray the evolution of weaknesses on a timeline of blocks.
We look at the percentage of contracts flagged by a particular tool as possessing any weakness (\autoref{fig:findings}) as well as at the percentage of contracts flagged by any tool as possessing a particular weakness (\autoref{fig:swc_classes}).

\subsection{Tool reports}\label{sec:tool_reports}
\autoref{fig:findings} depicts the reporting rate of each tool over the range of 14\,M blocks.
Each data point represents the percentage of bytecodes in a bin of 100\,k blocks that were marked with at least one non-omitted finding by the respective tool.
The vertical lines in gray indicate forks that added EVM opcodes and thus may affect weakness detection.
To improve readability, the diagram is split into three plots with four tools each.

\ITEM{Upper plot.}
Pakala (green) and teEther (red) both flag a few bytecodes only.
This can be attributed to the fact that Pakala scans for two rather infrequent weaknesses (SWC\,105, 112), and teEther just for one (SWC\,105).
eThor (orange) also scans for a single weakness (SWC\,107), albeit for a far more prevalent one.
Conkas (blue) scans for five weaknesses (SWC\,101, 104, 107, 114, 116), among them the most frequent ones (SWC\,101 and 107).

\ITEM{Middle plot.} 
MadMax (blue) and Ethainter (orange) specialize in rather specific weaknesses that they detect for a small number of contracts only.
MadMax is geared towards three gas issues, loosely related to SWC\,101, 113, and 128, while Ethainter scans for five weaknesses (SWC\,105, 106, 112, 124, and unchecked tainted static call).
Securify (red) also scans for five weaknesses (SWC\,104, 105, 107, 114, and missing input validation), including the popular reentrancy bug.

Mythril (green) tests for the largest number of weaknesses (SWC\,101, 104--107, 110, 112, 113, 115, 116, 120, 124, 127), including the most prevalent ones.
While we see a peak with more than 90\,\% of contracts flagged in the early days, the rate of contracts with reported weaknesses continuously drops to below 40\,\% towards the end of the timeline.

\ITEM{Lower plot.}
The tools in this plot differ from the others, as the rate of flagged contracts stays high or even increases towards the end of the timeline.
Maian (blue) scans for three weaknesses (SWC\,105, 106, and locked Ether), Osiris (orange) for nine (SWC\,101, 107, 114, 116, integer issues beyond SWC\,101, and the callstack depth bug), Oyente (green) for four (SWC\,107, 114, 166, and the callstack depth bug), and Vandal (red) for five (SWC\,104--107 and 115).

In terms of EVM operations supported (see \autoref{sec:supported_ops}), Maian, Oyente and Osiris are the oldest tools in our collection.
They do not handle the operation \OP{SHR}, which is central to newer contracts (\autoref{fig:ops_evol}).
Hence, we expect the rate of findings to drop over time rather than to rise.
It turns out that Oyente checks for \emph{Callstack Depth Bugs} by searching for a specific code pattern (instead of using symbolic execution as for the other weaknesses), and Osiris inherits this functionality from Oyente.
Even though the bug has become obsolete with the fork at block 2.463\,M (see \autoref{sec:synopsis_weaknesses}), the pattern is detected at an increasing rate and causes spurious findings.
Regarding Maian, it checks, among other weaknesses, for \emph{Ether lock}.
This property requires to check all execution paths for the absence of operations that are able to transfer Ether.
As the inability to handle \OP{SHR} cuts short more and more of the paths, the number of falsely reported Ether locks increases.

Vandal reports \np[\%]{96.6} of the contracts with a \OP{CALL} instruction as containing an \textit{Unchecked Call} and \np[\%]{88.4} as containing a \textit{Reentrant Call}.
Given that the majority of calls are method calls, for which the Solidity compiler adds checks automatically, and given that reentrancy is known to be a common but not a universal problem, we suspect that Vandal applies weak criteria and thus reports numerous false positives.

In \autoref{fig:findings2}, we omit the problematic findings \emph{Callstack Depth Bug} for Oyente and Osiris, the finding \emph{Ether lock} for Maian, and the weaknesses \textit{unchecked call} (SWC\,104) and \textit{reentrant call} (SWC\,107) for Vandal.
With these omissions, the number of flagged contracts either is constantly low or drops low.

\ITEM{General Observation.}
Overall, the share of flagged contracts diminishes over time.
For unmaintained tools, this may be related to the fact that they are no longer able to analyze recent contracts containing e.g.\ new instructions.
Moreover, code patterns tailored to specific compiler versions may fail to detect a weakness in bytecode obtained by later versions.
For actively maintained or new tools, the decreasing number of flagged contracts may indeed indicate that newer contracts are less vulnerable than older ones.

\subsection{SWC classes detected}

For weaknesses mapped to a suitable SWC class, \autoref{tab:vuls} gives an overview of their prevalence.
The column \emph{frequency} counts the number of unique skeleton bytecodes, where at least one tool reports the respective weakness%
\footnote{%
  Most tools do not verify their assessment by providing an exploit (like teEther does) or by proving the absence of the vulnerability (like eThor does).
  Hence, the table counts warnings rather than vulnerabilities.}.
As the tools tackle differing subsets of the SWC classes, the number of tools addressing a specific weakness varies from one to seven.
Due to our cumulative counting, the frequency of a weakness increases with the number of tools claiming to detect it, especially with overreporting tools.

\autoref{fig:swc_classes} depicts the 15 SWC classes on the timeline of 14\,M blocks.
For every SWC class, a data point represents the percentage of skeleton bytecodes in a bin of 100\,k blocks that were marked with the respective weakness by at least one tool.
The top plot shows the classes detected by four or more tools (SWC\,101, 105, 107, 114, 116), the middle one those handled by two or three tools (SWC\,104, 106, 112, 113, 124), and the third one those addressed by just one (SWC\,110, 115, 120, 127, 128).
In accordance with our discussion of Vandal above, we omit its findings from the plots, as its excessive reporting for SWC\,104 and 107 would distort the picture.

We see five weaknesses decrease over time from a high ($\geq$~\np[\%]{50}) or medium (\np[\%]{20}) level to a medium or low ($\leq$~\np[\%]{10}) level:
The findings of classes 101, 104, 107, 110, and 114 start to fall from about block~4\,M onwards.
The other 10 weaknesses stay on a steady, but low level after block~4\,M, except for 113 (middle plot), which fluctuates around \np[\%]{10} and 116 (top plot), which fluctuates around \np[\%]{20}.

The decline of potential integer overflows (101) seems plausible: Since version 0.8.0, the Solidity compiler adds appropriate checks automatically, and already some time before, the use of math libraries with the same effect had become quasi-standard.
Reentrancy (107) is probably the most (in)famous vulnerability.
The decrease in detection can be attributed at least partially to developers taking adequate precautions.

\newpage
\begin{mdframed}[style=mpdframe]
  \textbf{Observation 2.}
  Of the 37 SWC classes, 15 are covered by at least one tool, and 7 by at least three tools.
  For all weaknesses, the number of flagged contracts decreases over time or stagnates on a low level.
  The decreasing detection rates can be attributed to unmaintained tools that do not adequately cope with newer EVM instructions as well as to compilers and programmers taking counter-measures.
  At the end of the timeline, \emph{integer bugs} (SWC\,101), \emph{reentrancy} (SWC\,107) and \emph{block values as a proxy for time} (SWC\,116) are the most frequently detected weaknesses with a share of about 20\,\% each.
\end{mdframed}
\newpage
\section{RQ3 Tool Quality over Time} \label{sec:rq3}
To assess the quality of the tools, we consider execution times as well as their errors and failures.

\begin{table*}
\caption{Average Runtimes of Tools}
\label{tab:runtimes}
\centering
\begin{tabular}{lrrrrr}
\toprule
Tool       &Overall                         &Success    &Error       &   OOM      & Prg.issues\\
\midrule
Conkas	   &                     \np[s]{119}& \np[s]{84}&  \np[s]{28}& \np[s]{719}& \np[s]{21}\\
Ethainter  &\cellcolor{greenlight}\np[s]{71}& \np[s]{20}&\np[s]{1753}&            &           \\
eThor	   &\cellcolor{redmedium}\np[s]{574}& \np[s]{90}&  \np[s]{17}& \np[s]{874}& \np[s]{10}\\
MadMax	   &\cellcolor{greendark} \np[s]{21}& \np[s]{21}&\np[s]{1752}&            &           \\
Maian	   &\cellcolor{greenlight}\np[s]{48}& \np[s]{58}&  \np[s]{18}&\np[s]{1104}& \np[s]{62}\\
Mythril	   &\cellcolor{redmedium}\np[s]{670}&\np[s]{660}&            & \np[s]{834}&\np[s]{129}\\
Osiris     &                     \np[s]{165}&\np[s]{165}& \np[s]{165}&\np[s]{1242}&\np[s]{290}\\
Oyente	   &\cellcolor{greenlight}\np[s]{35}& \np[s]{34}&            & \np[s]{171}& \np[s]{30}\\
Pakala	   &\cellcolor{reddark} \np[s]{1115}&\np[s]{785}&   \np[s]{4}&\np[s]{1154}&\np[s]{627}\\
Securify   &                     \np[s]{160}&\np[s]{122}&  \np[s]{19}&            &\np[s]{940}\\
teEther	   &\cellcolor{redmedium}\np[s]{572}& \np[s]{72}& \np[s]{754}& \np[s]{886}&\np[s]{441}\\
Vandal	   &\cellcolor{greenlight}\np[s]{63}& \np[s]{41}&            & \np[s]{665}& \np[s]{26}\\
\bottomrule
\end{tabular}
\end{table*}

\ITEM{Execution time}.
\autoref{tab:runtimes} gives the average runtimes in seconds for each tool.
The column \emph{Overall} averages over all \np{248328} runs, whereas \emph{Success} picks only those completing without errors and failures.
The column \emph{Error} shows the average time for runs where the tool reports an error, while \emph{OOM} collects the runs terminated by an out-of-memory exception.
The last column, \emph{Prg.issues}, averages over runs with programming issues, like exceptions caused by type errors.
The average time for runs timing out is not listed explicitly, as it is close to \np[s]{1800} (\np[m]{30}), for obvious reasons.

Overall, the fastest tools are MadMax, Oyente, Maian, Vandal, Ethainter.
The slowest one, by far, is Pakala, with the next ones, Mythril, eThor, and teEther, being twice as fast. 
When considering only runs without errors and failures, eThor and teEther are substantially faster than on average, while Pakala and Mythril are still slow.
Mythril, Oyente, and Vandal do not report any errors, hence no times are listed in the respective column.
The average times on error are small for Pakala, Securify, Maian, eThor and Conkas, which indicates that most reported errors are show-stoppers.
For Madmax and Ethainter, the few errors are related to a timeout, hence the average is high.

\begin{figure}
\centering
\includegraphics[width=.9\columnwidth]{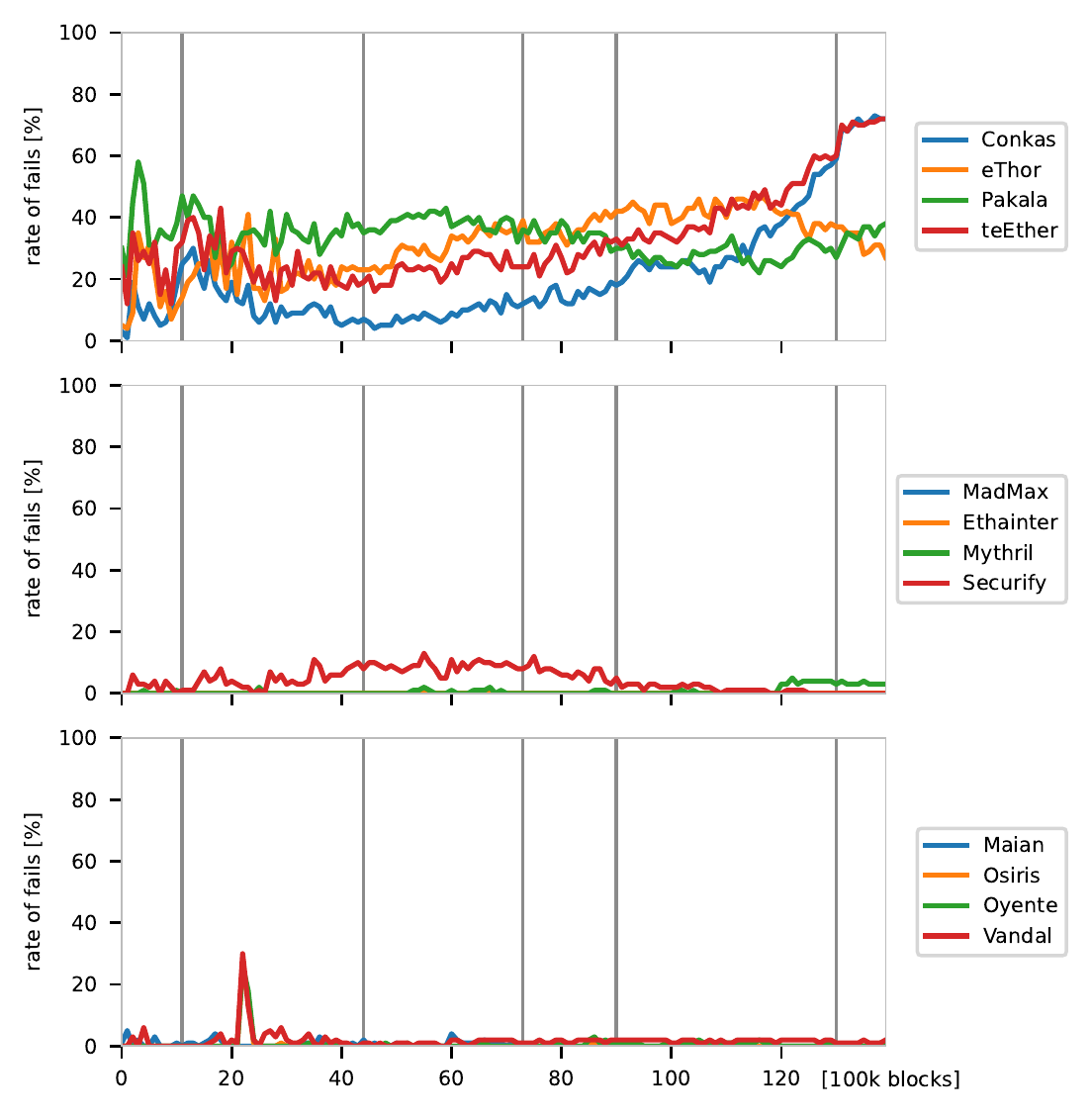}
\caption{Tool failures over time. Each data point shows the percentage of failures encountered by the tools, in bins of 100k blocks. Ethainter and MadMax had no failures.}
\label{fig:failures}
\end{figure}

\ITEM{Errors and Failures.}
We consider a run failed if it is terminated by an external timeout, an out-of-memory exception, or a tool-specific unhandled condition.
A run terminates properly if it stops under control of the tool, either successfully or with an error condition detected by the tool.
\autoref{fig:failures} depicts the failures over time in bins of 100\,k blocks as percentage of bytecodes where a tool fails. 
Conkas, eThor, Pakala, and teEther fail most often and are shown at the top, while the other tools show few or no failures.

\begin{table*}[!ht]
\caption{Findings, Errors and Failures of Tools}
\label{tab:finds}
\centering
\newcommand\HL{\cellcolor{reddark}}
\makebox[0pt]{%
\begin{tabular}{l|r@{~~}r|r@{~~}r|rrrr}
\toprule
           & \multicolumn{4}{c|}{Bytecodes with/without findings} & \multicolumn{4}{c}{Bytecodes with errors/failures } \\
Tool       & \multicolumn{1}{c@{~~}}{with} & +err/fail & \multicolumn{1}{c@{~~}}{without} & +err/fail& Error & Timeout & OOM  & Prg.issues \\
\midrule
Conkas     &\np{121436}& \np[\%]{0}&\np{126892}&\np[\%]{69}&    \np{16790}&    \np{7825}&      \np{44}&\HL\np{63111}\\
Ethainter  & \np{17842}& \np[\%]{0}&\np{230486}& \np[\%]{3}&     \np{7290}&       \np{0}&       \np{0}&       \np{0}\\
eThor      &\np{148643}& \np[\%]{0}& \np{99685}&\np[\%]{99}&     \np{9892}&\HL\np{71290}&     \np{369}&\HL\np{17407}\\
MadMax     & \np{11868}& \np[\%]{0}&\np{236460}& \np[\%]{0}&       \np{13}&       \np{0}&       \np{0}&       \np{0}\\
Maian      & \np{31929}&\np[\%]{84}&\np{216399}&\np[\%]{41}&\HL\np{114826}&    \np{1298}&      \np{28}&      \np{10}\\
Mythril    &\np{135818}& \np[\%]{0}&\np{112510}& \np[\%]{3}&        \np{0}&    \np{2620}&      \np{18}&    \np{1022}\\
Osiris     &\np{122883}&\np[\%]{53}&\np{125445}&\np[\%]{69}&\HL\np{151010}&      \np{86}&      \np{14}&     \np{576}\\
Oyente     & \np{67752}& \np[\%]{0}&\np{180576}& \np[\%]{1}&        \np{0}&      \np{11}&    \np{1929}&     \np{563}\\
Pakala     &  \np{4232}&\np[\%]{46}&\np{244096}&\np[\%]{34}&       \np{10}&\HL\np{80808}&     \np{333}&    \np{2574}\\
Securify   & \np{40699}& \np[\%]{3}&\np{207629}&\np[\%]{10}&    \np{10525}&    \np{1651}&       \np{0}&    \np{9586}\\
teEther    &  \np{3230}& \np[\%]{3}&\np{245098}&\np[\%]{41}&     \np{3752}&\HL\np{52250}&\HL\np{40306}&    \np{6608}\\
Vandal     &\np{187506}& \np[\%]{0}& \np{60822}& \np[\%]{8}&        \np{0}&    \np{2662}&    \np{1142}&    \np{1047}\\
\bottomrule
\end{tabular}}
\end{table*}

\autoref{tab:finds} gives an overview of the accumulated errors and failures by category.
The left part lists the number of bytecodes with and without finding, as well as the share accompanied by an error or failure.
In its right part, the table gives the number of bytecodes, where the analysis resulted in an error message and/or a failure due to a timeout, an out-of-memory condition or a program issue.

\begin{figure}
\centering
\includegraphics[width=.9\columnwidth]{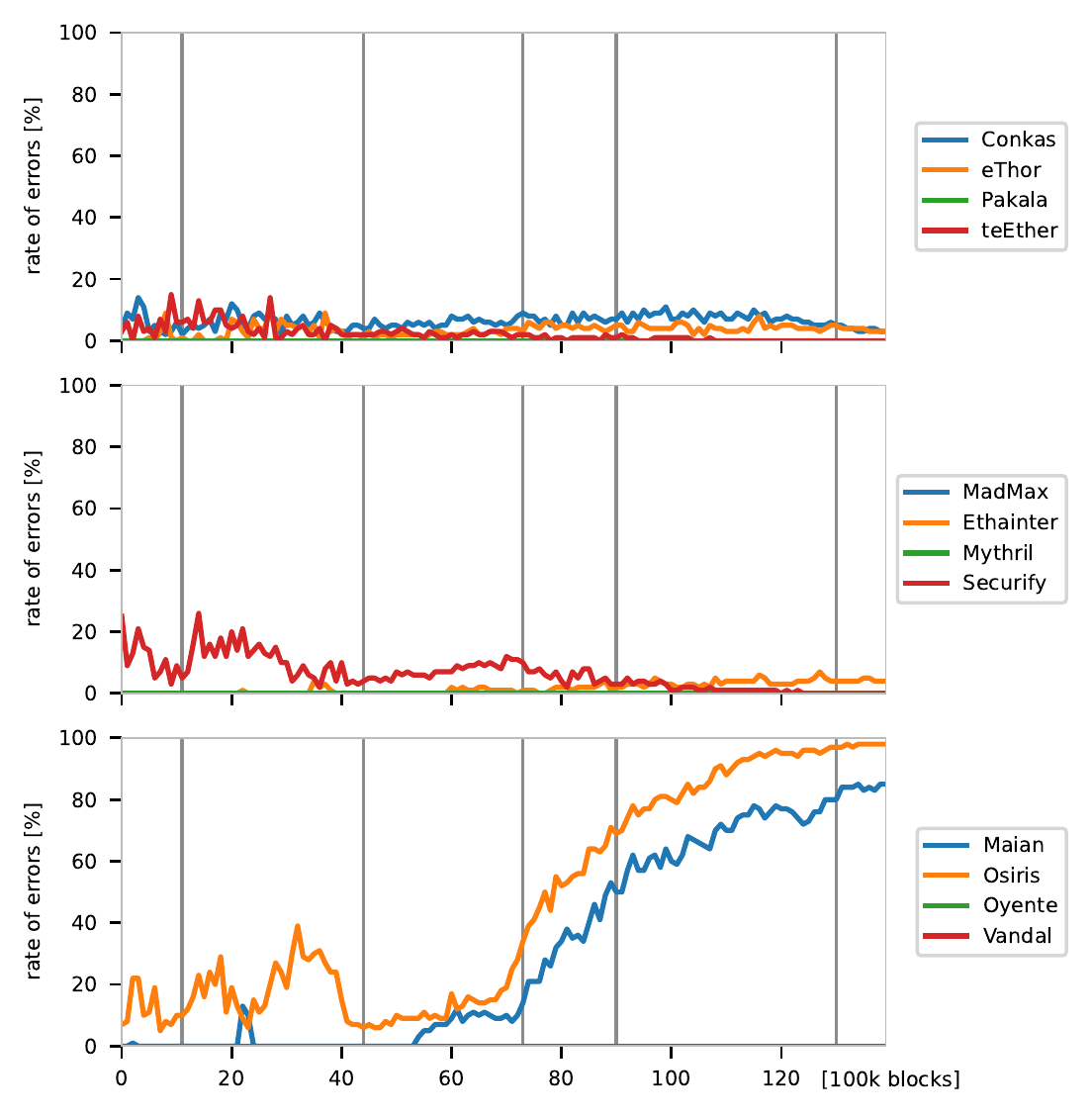}
\caption{Tool errors over time. Each data point shows the percentage of errors reported by the tools, in bins of 100k blocks. 
Mythril, Oyente and Vandal had no errors.}
\label{fig:errors}
\end{figure}

While most reported findings are not accompanied by any errors or failures, there are three notable exceptions.
Maian detects numerous occurrences of \emph{Ether lock} in spite of encountering unknown instructions.
The same accounts for Osiris when it reports the \emph{Callstack bug}.
This is due to the fact that the tools apply local pattern matching instead of symbolic execution.
Pakala reports a timeout for almost half of its analyses with findings.

eThor, Pakala and teEther show a large number of timeouts (marked red in \autoref{tab:finds}), which results in high average runtimes (marked red in \autoref{tab:runtimes}).
While Mythril shows a similarly high average runtime, it only has a low number of timeouts.
In contrast to the other three tools, it offers a parameter for getting notified about the external timeout and so is able to finish in time.

Regarding out-of-memory exceptions, only teEther sticks out.
Even with \np[GB]{32} of memory, it still fails for \np[\%]{16} of the inputs.

The last column in \autoref{tab:finds}, program issues, indicates to some extent the maturity of the tools' code.
Conkas fails for \np{63111} runs, with the most common causes being \emph{maximum recursion depth exceeded} \percent[b]{55626}{63111}; assertion failures \percent[b]{2499}{63111}; and type errors \percent[b]{2038}{63111}.
The \np{17407} fails of eThor result from the instruction EXTCODEHASH not being processed properly \percent[b]{11366}{17407}, arithmetic exceptions \percent[b]{5930}{17407}, and runtime exceptions \percent[b]{111}{17407}.
Securify fails for \np{9586} runs, mainly because of \emph{null pointer} exceptions \percent[b]{9496}{9586}.
Mythril, as the only tool actively maintained according to the activity on Github, fails for only \np{1022} bytecodes, the most frequent cause being type errors \percent[b]{952}{1022}, predominantly due to undefined terms in integer expressions.
At the lower end, we find Ethainter and MadMax with no program issues at all, and Maian, Osiris and Oyente with just a few.

\begin{mdframed}[style=mpdframe]
  \textbf{Observation 3.}
  Regarding resource consumption, a few tools require less than \np[s]{60} per contract with just a few GB of memory, whereas others  regularly approach the limits of \np[min]{30} and \np[GB]{32}.
  The rate of tool-reported errors varies between \np[\%]{0} and \np[\%]{60}, with the high rates resulting from tools operating outside of their specification.
  Questionably, there are tools with similar limitations but without any error at all.
  Regarding robustness, eight tools throw an exception for less than \np[\%]{1} of the contracts, as opposed to one tool with \np[\%]{25} fails.
  Program issues like type exceptions may be a consequence of using the dynamically typed language Python.
\end{mdframed}
\newpage
\section{RQ4 Overlap Analysis} \label{sec:rq4}
In this section, we investigate to which extent the tools agree in their judgments.
We use the SWC registry as a common frame of reference and map all findings to an appropriate SWC class, if any.

This excludes findings that do not fit any SWC class.
More specifically, the following nine 
findings are omitted for that reason: 
one finding of Ethainter (\emph{unchecked\_tainted\_static call}), 
one of Securify (\emph{missing\_input\_validation}),
one 
of Maian (\emph{Ether\_lock}
), 
five of Osiris (\emph{Callstack\_bug}, \emph{Division\_bugs}, \emph{Modulo\_bugs}, \emph{Signedness\_bugs}, \emph{Truncation\_bugs}), and one of Oyente \linebreak(\emph{Callstack\_Depth\_Attack\_Vulnerability}).

\newcommand\flagged{\mathrm{Flagged}}%
\newcommand\swc{\mathrm{Swc}}%
\newcommand\overlap{\mathrm{Overlap}}%
To determine the degree of overlap, we use the following measure.
For a tool~$t$, let $\swc(t)$ be the set of SWC classes that $t$ is able to detect, and let $\flagged(t,s)$ be the set of contracts that $t$ flags for having a weakness of class~$s$.
We define the overlap between two tools $t_1$ and $t_2$ as
\[ \overlap(t_1,t_2) = \frac
{\sum_{s\in\swc(t_1)\cap\swc(t_2)}\lvert\flagged(t_1,s)\cap\flagged(t_2,s)\rvert}
  {\sum_{s\in\swc(t_1)\cap\swc(t_2)}\lvert\flagged(t_1,s)\rvert}
\]
The numerator counts, per weakness, the contracts flagged by both tools, while the denominator gives the number of all contracts flagged by the first tool.
This measure is not symmetric. $\overlap(t_1,t_2)=100\,\%$ means that for the SWC classes in common, $t_1$ flags a subset of the contracts flagged by~$t_2$.
If additionally $\overlap(t_2,t_1)=100\,\%$ holds, then the two tools are in perfect agreement, something to be expected for $t_1=t_2$ only.

\begin{table}
  \caption{Overlap of Tool Findings [\%]}
  \label{tab:agreement}
  \centering
  \renewcommand\tabcolsep{1pt}
\def\cca#1{%
  \ifx\relax#1\relax\else
    \cellcolor{DarkGreen!#1}%
    \ifdim #1pt>49pt\color{white}\fi
    #1%
  \fi}
  \makebox[0pt]{\small
  \begin{filecontents*}{overlaptable.csv}
Tool;Conkas;Ethainter;eThor;MadMax;Maian;Mythril;Osiris;Oyente;Pakala;Securify;teEther;Vandal
Conkas;100.0;;56.8;1.6;;28.7;26.2;6.7;;14.8;;85.1
Ethainter;;100.0;;;14.3;15.6;;;17.1;4.7;13.0;31.4
eThor;68.3;;100.0;;;22.2;5.6;2.3;;6.6;;76.0
MadMax;39.2;;;100.0;;24.9;46.6;;;;;
Maian;;38.4;;;100.0;65.3;;;57.6;17.9;81.5;44.9
Mythril;52.3;35.9;49.6;1.5;28.9;100.0;22.3;5.8;60.0;9.0;34.9;85.7
Osiris;58.4;;49.8;2.8;;35.6;100.0;58.5;;44.0;;94.5
Oyente;35.1;;54.7;;;55.2;90.2;100.0;;50.0;;93.5
Pakala;;24.6;;;29.4;61.6;;;100.0;9.0;44.9;23.8
Securify;38.1;13.2;50.6;;21.0;17.7;41.0;29.8;20.6;100.0;27.3;70.9
teEther;;19.3;;;50.2;43.2;;;54.2;14.4;100.0;30.7
Vandal;24.1;21.2;33.4;;11.3;11.4;4.7;1.7;11.0;4.0;11.8;100.0
\end{filecontents*}
\pgfplotstabletypeset[
    col sep=semicolon,
    precision=1,
    fixed zerofill,
    every head row/.style={before row=\toprule, after row=\midrule},
    every last row/.style={after row=\bottomrule},
    tool/.style={column type=r, assign cell content/.style={@cell content=\cca{##1}}},
    tool/.list={1,2,3,4,5,6,7,8,9,10,11,12},
    columns/Tool/.style={string type, column type=l, column name=$t_1{\downarrow}$\hskip 0pt plus 1filll$t_2{\rightarrow}$},
]{overlaptable.csv}}
\end{table}

\autoref{tab:agreement} shows the overlap between any two tools, with $t_1$ listed vertically and $t_2$ horizontally.
Since eThor detects reentrancy only, its row and column in the table give an idea of how differently a weakness may be assessed by the tools.
For Vandal, we find high values in its column and low ones in its row, which indicates that most weaknesses it reports are not backed by other tools.
As discussed in \autoref{sec:tool_reports}, a large number of Vandal's findings are likely to be false positives, and the numbers in \autoref{tab:agreement} reflect that.

Another observation concerns Osiris and Oyente.
We expect a high overlap as Osiris extends Oyente.
In fact, \np[\%]{90.2} of Oyente's findings are backed by Osiris, while Oyente covers \np[\%]{58.5} of Osiris' findings.
Apparently, Osiris not only detects additional weaknesses (not considered in the comparison), but also flags additional contracts with weaknesses the tools have in common.

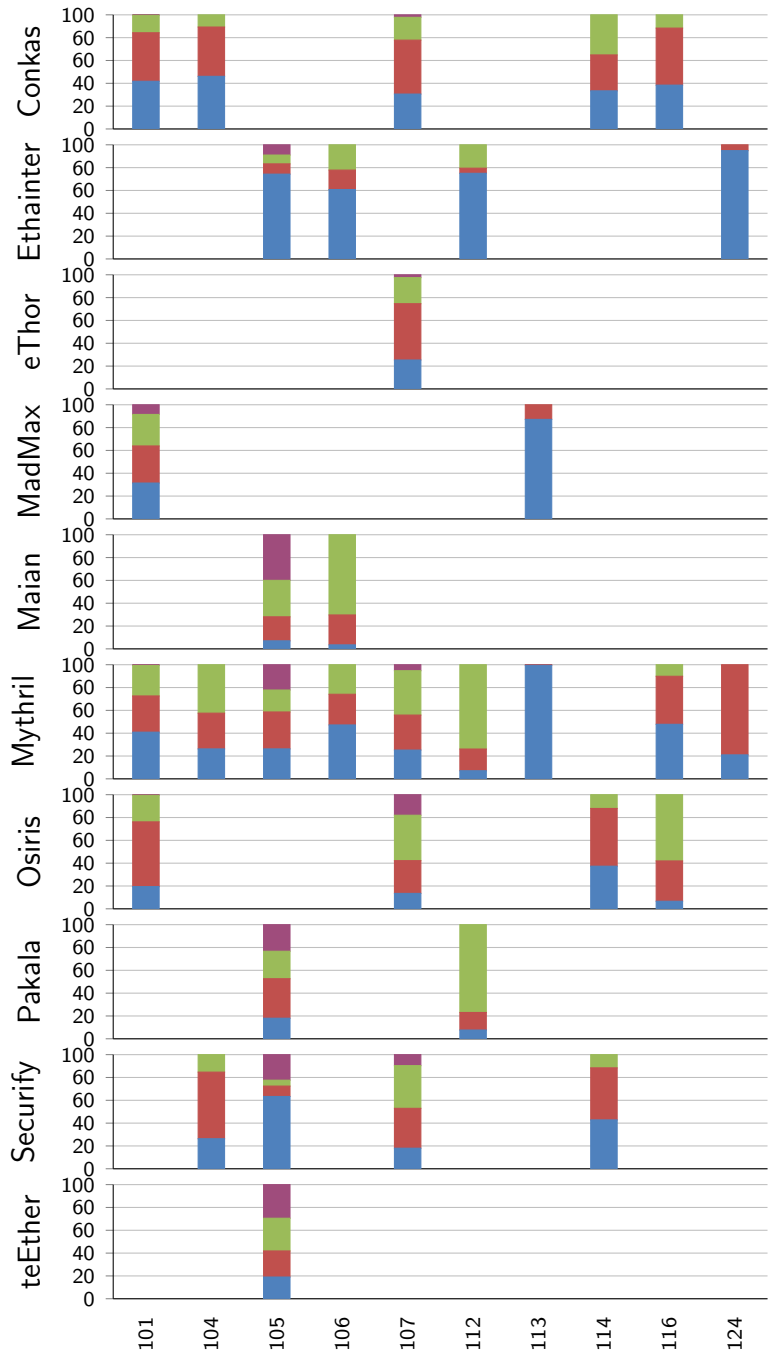
\begin{figure}
  \tikzset{declare function = { xpos(\x) =
    \x==101?0:(\x==104?1:(\x==105?2:(\x==106?3:(\x==107?4:(
    \x==112?5:(\x==113?6:(\x==114?7:(\x==116?8:(\x==124?9:nan)))))))));}}
  \newcommand\addlayer[2]{
    \addplot[style={#2,fill=#2,mark=none}] table[
      x expr={xpos(\thisrow{swc})},
      y expr={\thisrow{#2}/\thisrow{flagged}*100},
      col sep=comma
    ] {figure_agreement_#1.csv};\label{style:#1:#2}
  }
  \newcommand\addtool[2]{
    \nextgroupplot[ylabel = #2]
    \addlayer{#1}{one}
    \addlayer{#1}{two}
    \addlayer{#1}{three}
    \addlayer{#1}{more}
  }
  \centering
  \begin{tikzpicture}[font={\sffamily}]
    \begin{groupplot}[
      group style={
          group name=agreement_analysis,
          group size=1 by 10,
          xlabels at=edge bottom,
          xticklabels at=edge bottom,
          vertical sep=6pt
      },
      ybar stacked,
      footnotesize,
      width= 0.75*\textwidth,
      height=3.1cm,
      xmin=-0.5, xmax=9.5,
      ymin=0, ymax=100,
      unbounded coords=jump,
      major x tick style = transparent,
      ymajorgrids = true,
      scaled ticks=false,
      xtick={0,1,...,9},
      xticklabels={101, 104, 105, 106, 107, 112, 113, 114, 116, 124},
      x tick label style={rotate=90,anchor=east,font={\sffamily\scriptsize}},
      ytick={0,20,...,100},
      yticklabels={0, 20, 40, 60, 60, 100},
      y tick label style={font={\sffamily\scriptsize}},
      axis x line*=bottom,
      axis y line*=left,
      ylabel near ticks, 
      tickpos=left,
      ytick align=outside,
      xtick align=outside,
      legend style={
          at={(0.5,1.25)},anchor=north,
          legend columns=-1,
          draw=none,
          fill=none
      }
    ]
    \addtool{conkas}{Conkas}
    \addtool{ethainter}{Ethainter}
    \addtool{ethor}{eThor}
    \addtool{madmax}{MadMax}
    \addtool{maian}{Maian}
    \addtool{mythril}{Mythril}
    \addtool{osiris}{Osiris}
    \addtool{pakala}{Pakala}
    \addtool{securify}{Securify}
    \addtool{teether}{teEther}
    \end{groupplot}
  \end{tikzpicture}  
  \caption{Agreement of the tools' judgment on the SWC classes. Each bar shows the proportion of weaknesses identified by one \ref*{style:mythril:one}, two \ref*{style:mythril:two}, three \ref*{style:mythril:three}, and four or more \ref*{style:mythril:more} tools.}
  \label{fig:agreement}
\end{figure}

\autoref{fig:agreement} shows the overlap in more detail.
We exclude Vandal (due to its overreporting) and Oyente (as Osiris extends it), to avoid an inflation of overlaps.
Each row gives a breakdown of the contracts flagged by a specific tool, for each SWC class covered by at least two tools.
Blue identifies the share of contracts flagged exclusively by the tool, whereas red, green, and purple indicate the share also flagged by one, two, or more other tools.
A good agreement shows as purple where four or more tools check for the SWC class (101, 105, 107), green where three tools detect it (104, 106, 112, 114, 116), and red for two tools (113, 124).

\ITEM{SWC 101~-- \SWC{101}:}
We find hardly any agreement of all four tools. 
MadMax, by construction, checks for a subcase of 101 that is not covered by the other tools, but even green (overlap of three) is rare.
In \autoref{sec:relation}, we analyze the evolution of overlaps for this weakness in more detail.

\ITEM{SWC 104~-- \SWC{104}:}
The three tools show some agreement, as red and green dominate blue.

\ITEM{SWC 105~-- \SWC{105}:}
Detected by six tools, we see the highest amount of purple among all classes.

\ITEM{SWC 106~-- \SWC{106}:}
Virtually all of Maian's findings coincide with at least one other tool, while Ethainter and Mythril show a fair amount of blue.
The top plot of \autoref{fig:errors} provides an explanation: In the second half of the timeline, the error rate of Maian increases, as the tool fails to handle more recent contracts with new types of instructions, so Maian stops reporting weaknesses.

\ITEM{SWC 107~-- \SWC{107}:}
Even though reentrancy is one of the best-researched weaknesses and is detected by five tools, agreement of more than three tools is rare.
In \autoref{sec:relation}, we analyze the evolution of overlaps for this weakness in more detail.

\ITEM{SWC 112~-- \SWC{112}:}
This weakness is detected by three tools, hence the large amount of green actually indicates the best agreement in the chart.
Ethainter seems to implement a more liberal definition of the vulnerability, as it flags many additional contracts (blue).

\ITEM{SWC 113~-- \SWC{113}:}
MadMax has been designed to detect specific gas-related issues, which partly map to this class.
There is some overlap with Mythril, but since the latter flags many more contracts under this label, the red share is not visible in Mythril's bar.

\ITEM{SWC 114~-- \SWC{114}:}
The bars are mainly blue and red, indicating little agreement between all three tools.

\ITEM{SWC 116~-- \SWC{116}:}
Virtually all contracts flagged by Osiris are also flagged by one of the other tools, in most cases by both.
The other tools, however, flag many more contracts, as the comparatively small size of the green part~-- representing the same group of contracts in all three bars~-- shows.
Like in the case of Maian and SWC~106 above, the error rate of Osiris increases in the second half of the study period, as new instructions prevent it from reporting weaknesses (\autoref{fig:errors}).

\ITEM{SWC 124~-- \SWC{124}:}
The contracts flagged by Mythril are essentially a subset of those flagged by Ethainter, but a small one, as the blue part of Ethainter's bar dominates.

\begin{mdframed}[style=mpdframe]
\textbf{Observation 4.}
There is little agreement between the tools regarding the findings, even for well-researched and frequently analyzed weaknesses such as reentrancy.
Contributing factors are the lack of commonly accepted, precise definitions for the weaknesses as well as diverging approaches to detect them.
A mutually low agreement suggests that the tools are rather complementary.
\end{mdframed}
\newpage
\section{Discussion} \label{sec:discussion}

In this section, we combine the results of our research questions and discuss them in a wider context.

\subsection{Relation between Findings, Errors, Failures, and Overlap} \label{sec:relation}

In the last section, we looked at the overlap of tools, accumulated over time as well as over common SWC classes.
Here, we pick two exemplary SWC classes, take a closer look at the evolution of findings over time, and correlate the overlap of tools with their errors and failures.

\begin{figure}
\centering
\includegraphics[width=.78\columnwidth]{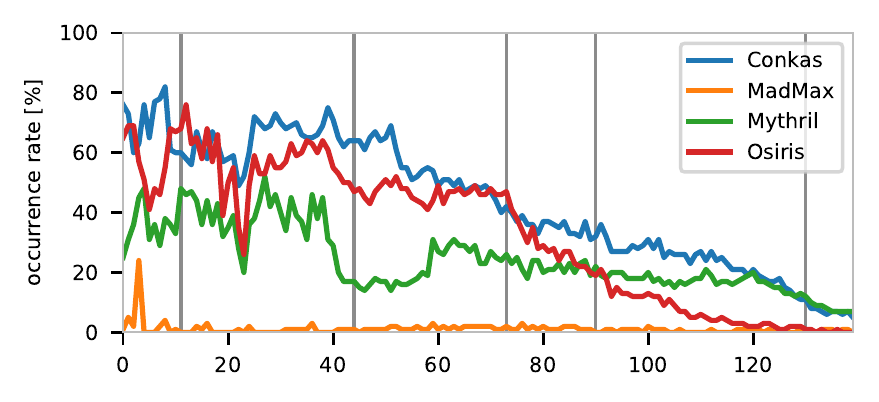}\\
\hspace*{15.5mm}%
\includegraphics[width=.79\columnwidth]{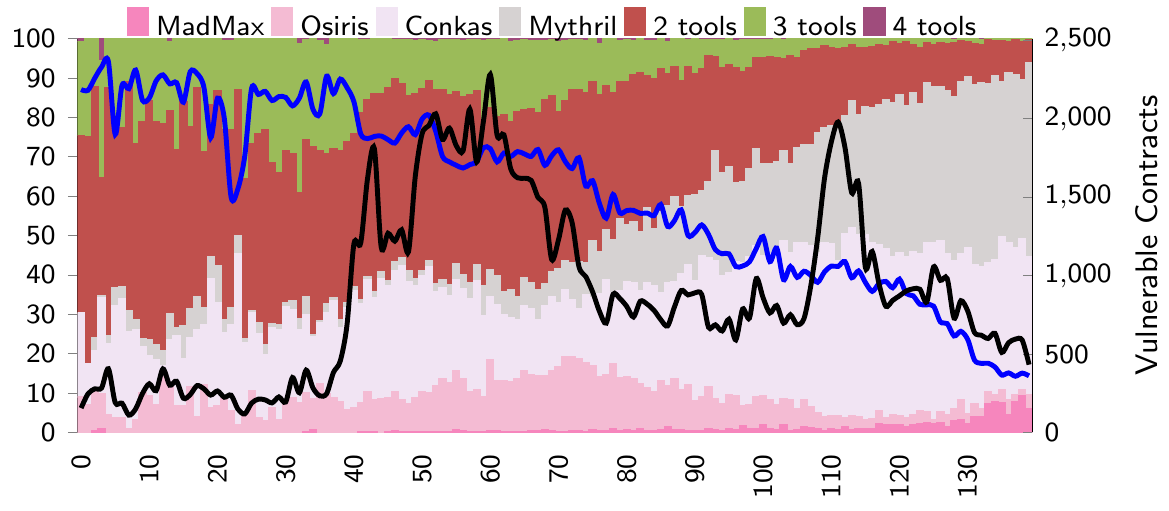}\\
\includegraphics[width=.78\columnwidth]{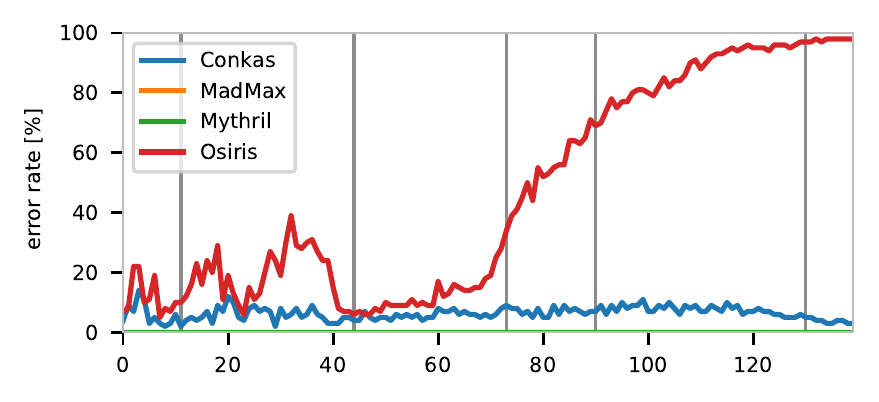}\\
\includegraphics[width=.78\columnwidth]{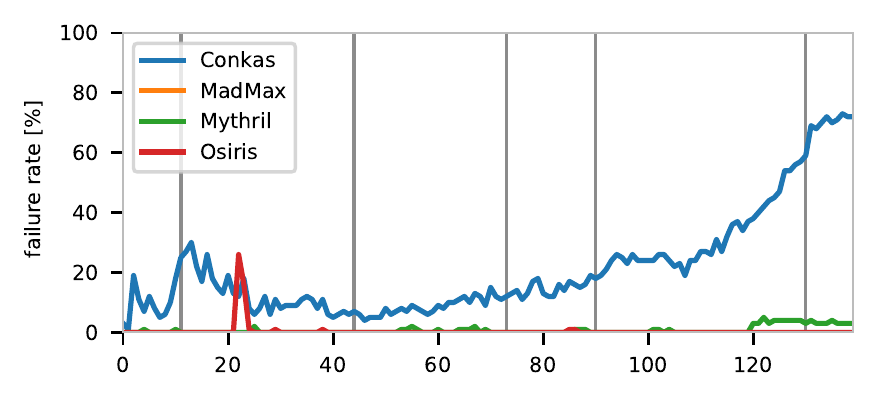}
\caption{SWC-101 Integer Overflow and Underflow on a timeline of blocks, in bins of \np[k]{100} blocks.
  Top: Percentage of bytecodes flagged, per tool.
  Upper middle: Percentage of overlaps. 
  Lower middle: Error rate of tools.
  Bottom: Failure rate of tools.}
\label{fig:swc101}
\end{figure}

\ITEM{SWC\,101~-- Integer Over- and Underflow.}
In \autoref{fig:swc101}, the top most plot depicts the percentage of bytecodes flagged with an integer over- or underflow, per tool.
Starting from different levels around \np[\%]{70} and \np[\%]{40}, Conkas and Mythril converge at \np[\%]{10} at the end of the timeline.
Osiris shows a weakness level comparable to these tools for most of the timeline, but falls to \np[\%]{0} towards the end. 
MadMax reports hardly any cases throughout the whole timeline.

The second plot in \autoref{fig:swc101} visualizes the agreement of tools over time.
The lines in the foreground show the number of contracts flagged by any tool, once in relative terms (blue line with the scale to the left, with 100\,\% corresponding to all contracts), and once in absolute terms (black line with the scale to the right, numbers per bin of 100\,k blocks).
The background divides the flagged contracts into shares that are flagged by a single tool, by two, three, or four tools, respectively.

Up to block 6\,M (bin 60), the brown and green areas with purple specks at the top show that 60\,\% of the flagged contracts are flagged by at least two tools.
The other 40\,\% are split between Osiris and Conkas, who are the sole tools flagging the respective contracts.
The gray area of contracts flagged solely by Mythril is small, even though the tool finds the weakness in 20--40\,\% of all contracts (top plot).
Apparently, at least one other tool agrees with Mythril most of the time.

The picture changes in the second half of the plot.
Towards the end of the timeline, there is hardly any agreement anymore.
Less than 10\,\% of the contracts are flagged by at least two tools, while most are flagged solely by Conkas or Mythril.

The situation can be partly explained by the fact that MadMax specializes in gas issues, with one of its findings constituting a specific type of overflow that occurs in a few contracts only (see top plot).
For Osiris, we see a rise in errors (third plot of \autoref{fig:swc101}) that mirrors the increased usage of the \OP{SHR} operation (\autoref{fig:ops_evol}), which is not supported by Osiris (\autoref{tab:operations}).
Therefore, the detection rate of Osiris drops to zero (top plot), leaving us essentially with two tools at the end of the timeline.
In spite of Conkas' failure rate rising to 70\,\% (fourth plot of \autoref{fig:swc101}), its detection rate remains comparable to Mythril.

From version 0.8.0 onwards, the Solidity compiler inserts checks for over- and underflows into the bytecode.
In view of the compiler's adoption rate (\autoref{fig:ops_evol}), it seems that the vulnerability has actually become extinct at the end of the timeline and that the respective findings of Mythril and Conkas are false positives.

\begin{figure}
\centering
\includegraphics[width=.78\columnwidth]{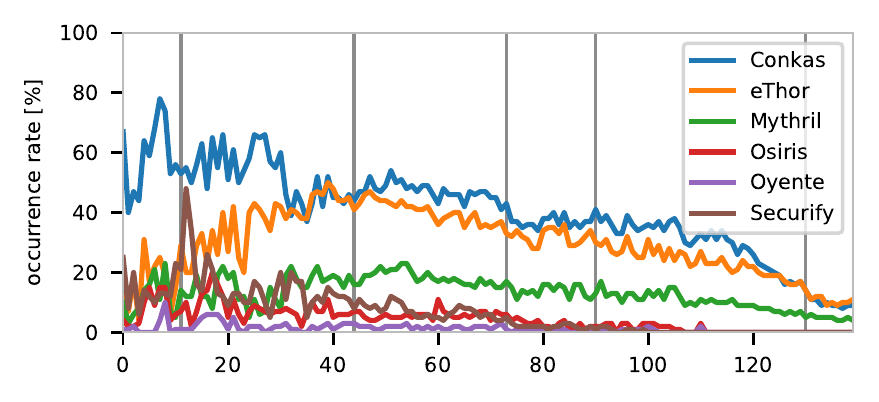}\\
\hspace*{15.5mm}%
\includegraphics[width=.79\columnwidth]{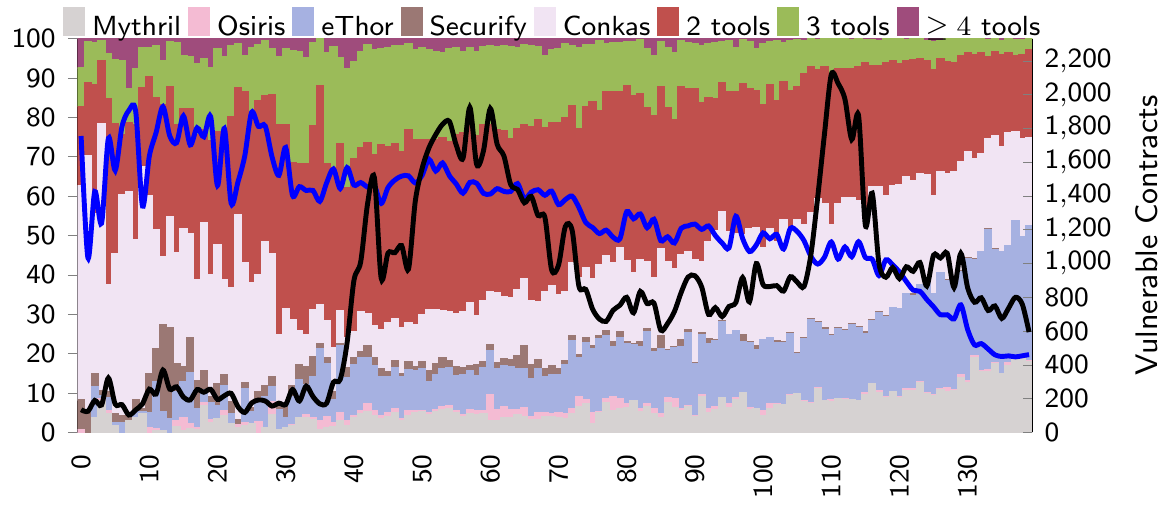}\\
\includegraphics[width=.78\columnwidth]{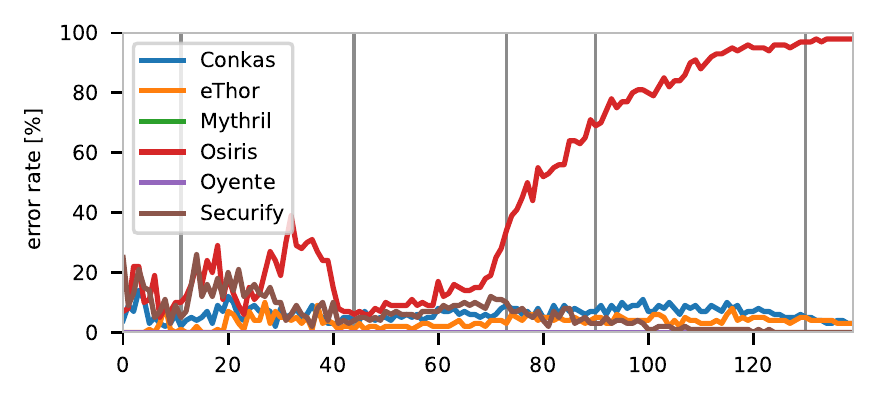}\\
\includegraphics[width=.78\columnwidth]{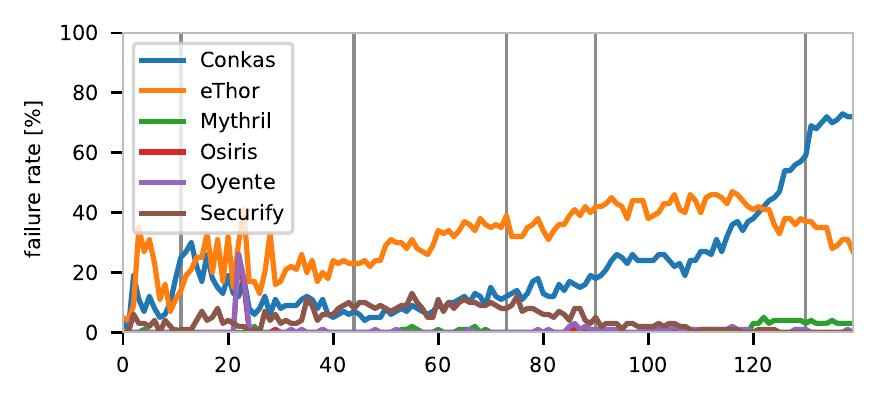}
\caption{SWC-107 Reentrancy on a timeline of blocks, in bins of \np[k]{100} blocks. 
  Top: Percentage of bytecodes flagged, per tool.
  Upper middle: Percentage of overlaps. 
  Lower middle: Error rate of tools.
  Bottom: Failure rate of tools.}
\label{fig:swc107}
\end{figure}

\ITEM{SWC\,107~-- Reentrancy.}
At a first glance, \autoref{fig:swc107} shows a situation similar to \autoref{fig:swc101}, just for another weakness and the six tools detecting it.
The detection rate of three tools (Osiris, Oyente, Securify) is low and drops to zero towards the end.
For Osiris and Oyente, the reason is again their inability to handle new operations, in particular \OP{SHR}, even though Oyente quits silently, while Osiris issues errors (third plot).
For Securify, the collected data does not provide an explanation for the diminishing detection rate.

Conkas and eThor exhibit significant failure rates (bottom plot in \autoref{fig:swc107}), but this does not prevent them from reporting up to 40\,\% of contracts as potentially vulnerable to a reentrancy attack.
On the surface, these two tools show a similar behavior, reporting similar rates of reentrant contracts from block \np[M]{3.5} (top plot, bin 35) onwards.
However, the Jaquard similarity for the flagged contracts (number of contracts flagged by both tools divided by the number of contracts flagged by at least one tool) is only \np[\%]{45} at block \np[M]{3.5}, and drops to \np[\%]{28} for the last part where the blue and orange lines seem to coincide.

This is also reflected in the second plot of \autoref{fig:swc107}, where the agreement of two or more tools (red, green and purple area) decreases steadily from block \np[M]{4.5} (bin 45) onwards, while the shares of contracts flagged exclusively by Conkas (pink), eThor (blue) or Mythril (gray) increase, such that at the end of the timeline, the four groups are roughly of the same size.

Our explanation for the disagreement between the tools for this weakness as well as for most others, is the lack of commonly agreed, unambiguous definitions, which is backed by our work on a unified ground truth~\cite{MdAGS2023GT}.
On the surface, the tools aim for the same weakness, motivated by similar examples, but the respective interpretations and implementations may differ considerably.

\subsection{Assessment of Tools}

Based on the results of our evaluation, we summarize the observed properties of the tools.

\ITEM{Conkas.}
With an average runtime of 119\,s and 4\,GB of memory, Conkas belongs to the light-weight tools.
The number of contracts timing out or running out of memory is small.
It seems that Conkas underwent a not entirely successful update for the operations of recent forks, as the source code seems to support them, but the tool fails on contracts using them, resulting in a high number of program exceptions.
This may be caused by a divergence between Rattle, the module generating the intermediate representation, and the analysis module on top.
Despite these problems, Conkas reports many findings.

\ITEM{Ethainter and MadMax.}
These two tools are among the most efficient and robust tools.
With average runtimes of 71\,s and 21\,s, respectively, and 4\,GB of memory, they are fast and never exceed the allotted memory.
The few errors reported are timeouts under control of the tools.
The number of failures is zero, indicating a high engineering quality.
This may be due to the robust base component Soufflé and the use of Datalog as a high-level specification language.

\ITEM{eThor.}
With an average runtime of 574\,s, a large number of timeouts, and 369 bytecodes running out of memory even with 32\,GB, this tool is one of the elephants in our study.
The high resource consumption may be caused by the complex workflow~-- eThor is the only tool trying to show the absence of a weakness.
The use of the strongly-typed programming language Java explains the absence of type errors (as we see with Python programs).
eThor seems to support the most frequent operations introduced by forks, but throws errors for some of them.
This makes the decreasing rate of reported reentrancy issues (and the increasing rate of contracts found secure) an unreliable indicator for the assumption that the frequency of reentrancy weaknesses indeed drops.

\ITEM{Maian.}
With an average runtime of 48\,s and 4\,GB memory sufficing for almost all bytecodes, Maian is a lightweight.
As it is the oldest tool and unmaintained, it supports hardly any of the newer operations.
This results in the second highest number of errors (reporting unknown opcodes) and virtually no weakness detections for newer contracts.

\ITEM{Mythril.}
With an average runtime of 670\,s, Mythril belongs to the slow tools, but almost never needs more than 4\,GB of memory.
It is the only actively maintained tool in our collection: Every issue we reported was fixed within a few days.
Mythril supports all EVM operations and checks for a large number of weaknesses.
This, and the tendency to report also issues of low severity, result in the third largest number of flagged contracts. 

\ITEM{Oyente and Osiris.}
With an average runtime of 35\,s and 165\,s, respectively, the two tools are among the faster tools.
Osiris extends Oyente and checks for further properties, which explains the additional time it takes.
4\,GB of memory suffice for most contracts.
However, Oyente runs out of 32\,GB of memory for about 2\,000 bytecodes, whereas Osiris seems to require less memory and hardly ever exceeds the quota.
Both tools fail for operations beyond fork 4.37\,M, with Osiris issuing a message and Oyente failing silently.
Consequently, both tools report no weaknesses for recent contracts.

\ITEM{Pakala.}
With an average runtime of 1115\,s, this tool is by far the slowest, which seems to be a consequence of Pakala performing symbolic execution without optimizations.
The tool author aimed at a small and simple program and deliberately omitted techniques like the construction of control flow graphs.%
\footnote{\url{https://www.palkeo.com/en/projets/ethereum/pakala.html}}
In spite of being able to handle all relevant operations, Pakala flags only 4\,232 contracts as vulnerable, which may have different causes.
First, the analysis of 80\,000 contracts timed out, so a prolonged analysis might have revealed further weaknesses.
Second, Pakala might actually spend the extra time for a more refined analysis, leading to a lower number of false positives.
Third, the simplicity of the program might have resulted in a lower detection rate.
To determine the actual cause, we would need to check the quality of the results, which is beyond the scope of our study.

\ITEM{Securify.}
The average runtime of 160s makes Securify one of the faster tools, even though it times out for 1651 bytecodes.
None of the runs exceeds the memory quota.
The implementation language Java prevents type errors, but we see almost 10\,000 null pointer exceptions.
Even though Securify has been superseded by a successor (that supports source code only and thus does not fall into the scope of our study) and is unmaintained now, it supports most essential EVM operations.
Nevertheless, its detection rate starts to drop early on and falls to virtually zero towards the end.

\ITEM{teEther.}
The average runtime of 572\,s is comparable to Mythril, but teEther times out in 52\,250 cases (compared to 2\,620 for Mythril).
The tool is exceptional regarding its appetite for memory: even with 32\,GB provided, 40\,306 analyses exceed the memory quota.
teEther addresses a single vulnerability that it detects in 3\,230 bytecodes.
The tool supports the essential EVM operations, but is unmaintained now.
With 6\,608 Python exceptions, teEther seems to be an experimental tool focusing on the elaborate analysis of a single issue.

\ITEM{Vandal.}
With an average runtime of 63\,s and 4\,GB of memory, Vandal is one of the fast and light tools, but still runs into a timeout for 2\,662 bytecodes and exceeds 32\,GB of memory for another 1\,142.
The number of 1\,047 programming issues is moderate for a tool written in Python.
With 75\,\% of the contracts flagged, Vandal surpasses the detection rate of the other tools.
The high rate triggered some plausibility checks in \autoref{sec:tool_reports}, showing that most bytecodes containing a \OP{CALL} operation are flagged as containing an unchecked or reentrant call.
Vandal seems to implement rather unspecific criteria that lead to a large number of false positives.
This interpretation is supported by Vandal's repository, where the patterns for weakness detection are listed just as use cases for a framework that decompiles bytecode to single static assignments.
The accompanying paper, on the other hand, presents Vandal as a tool for vulnerability detection.

\subsection{Combining or Comparing Tools Results}
When comparing or combining tool results, we face two challenges: (i) different aims of tools that are reflected in the way their findings are reported and (ii) differing definitions of weaknesses (that are associated with the findings), which makes it hard to map a finding to a class (within a common frame of reference) for comparison or combination. 

The tools can be divided into four groups with respect to their aim (for a specific weakness): (i) proving the absence of a property that is regarded as a weakness or vulnerability, (ii) over-reporting as to not overlook a potential weakness (aka issuing warnings), (iii) under-reporting since only those weaknesses are reported where a verification could be found (avoiding false alarms), (iv) reporting properties that are hardly a weakness (e.g.\ honeypots) or not necessarily (e.g.\ gas issues).

This distinction is important when comparing tools.
It strongly affects the number of agreements.
As we have seen in \autoref{sec:rq4} and \autoref{sec:relation}, the overall agreement is low, which is partly due to the fact that tools address different versions, subsets or supersets of a weakness class.
Considering the different aims of the tools, the low general agreement is not surprising.
However, it is even low for tools with similar aims.

The aims of the tools also impact voting schemes that combine the results of several tools to `determine' whether a contract is actually vulnerable. 
For over-reporting tools, it may make sense to have a majority vote.
However, under-reporting tools should rather be joined than intersected.

\subsection{Comparison to Source Code as Input}
We selected the tools in our study for their ability to process runtime code, as our goal was to analyze contracts deployed on the mainchain, for which Solidity source code is often unavailable.
Moreover, this allowed us to include Ethainter, eThor, MadMax, Pakala, teEther and Vandal, which require runtime code.
The other selected tools accept both, bytecode and Solidity source code.
In this section, we discuss the effect of using source code as the input.

Conkas, Osiris, Oyente and Securify compile the Solidity source to runtime code and then perform the same analysis as if the latter had been the input.
There are two differences, though.
First, the tools are able to report the location of weaknesses within the source, as they use a mapping provided by the compiler to translate bytecode addresses back to line numbers.
Second, for Solidity sources with more than one contract, the tools compile and analyze each one separately.
As complex contracts are structured into several layers of intermediate contracts using inheritance, this leads to redundant work.
While compilation and address mapping incur a negligible overhead, the additional contracts may lead to fewer or more findings within a fixed time budget, depending on whether there is less time for the main contract or whether other contracts contribute additional findings.\footnote{%
An easy remedy would be to extend the tools by a parameter with the name of the contract to analyze.}

Maian and Mythril compile the Solidity source as well but proceed with the deployment code, which includes contract initialization as well. 
Maian deploys the contract on a local chain and checks some properties live, like whether the contract accepts Ether. 
Moreover, the findings are filtered for false positives by trying to exploit the contract on the chain. 
Mythril, on the other hand, uses the deployment code to analyze also the constructor. 
For both tools, resource requirements and results will vary with the chosen form of input.

\subsection{Threats to Validity}
\label{sec:threats}

\ITEM{Internal validity} is threatened by integrating the new tools into SmartBugs.
We mitigated this threat by carefully following the SmartBugs instructions for tool integration and by consulting the documentation and the source code of the respective tools.
Multiple authors manually analyzed all execution errors to ensure that we had configured the tools adequately.
Moreover, we make the implementation and the results accessible for public inspection.

\ITEM{External validity} is threatened by the use of single bytecodes as proxies for code families identified by the same skeleton.
These representatives may not accurately reflect the code properties of all family members that are relevant to weakness detection.
We mitigated this threat by the first research question.
However, the random sample of \np{1000} bytecodes (\np{620} code families) may have been chosen too small such that our answer to RQ1 may not generalize to all bytecodes.

The focus on runtime bytecode as the sole object of analysis restricts the number of tools usable for our study, as well as the methods applicable.
Some trends and observations may thus not generalize to smart contract analysis in general.

\ITEM{Construct validity} is threatened by our mapping of the detected weaknesses to the classes of the SWC registry.
The mapping reflects our understanding of the weaknesses and what the tools actually detect, which may be incorrect.
We mitigated this risk by involving all authors during the mapping phase and by discussing disagreements until we reached a consensus.

Another potential threat are the resources, 30 minutes and up to 32\,GB per tool and bytecode.
This configuration is in line with related work or surpasses it.

\newpage
\section{Related Work}\label{sec:related}
 \begin{table}
 \caption{Overview of Evaluations with Benchmarks}
 \label{tab:related}
 \setlength{\tabcolsep}{2pt}
 \centering
 \newcolumntype{B}{>{\columncolor{midblue}}c}
     \makebox[0pt]{\footnotesize
     \begin{tabular}{@{}ll|cccBccBcBBB|r@{}}
 \toprule
Reference && \SW{\cite{Dika2017}} & \SW{\cite{Parizi2018empirical}} & \SW{\cite{Gupta2020}} & \SW{\cite{Durieux2020}} & \SW{\cite{ghaleb2020effective}} & \SW{\cite{Leid2020}} & \SW{\cite{Zhang2020a}} & \SW{\cite{Dias2021}} & \SW{\cite{Ji2021}} & \SW{\cite{Ren2021}} & \SW{\cite{Kushwaha2022tools}} & \SW{Our study} \\ 
 \midrule
 Evaluation Year && \multicolumn{1}{c|}{2017} & \multicolumn{1}{c|}{2018} & \multicolumn{2}{c|}{2019} & \multicolumn{3}{c|}{2020} & \multicolumn{4}{c|}{2021} & 2022\\ 
 \midrule
 \rowcolor{white}
 \multicolumn{2}{r|}{vulnerable}		& \cellcolor{bad}23 & \cellcolor{bad}10 & 162 & 69 & \cellcolor{good}\np{9369} & \cellcolor{bad}39 & 176  & 94 & 237 & 214 & ~ \\ 
 \rowcolor{white}
 \multicolumn{2}{r|}{non-vulnerable} & \cellcolor{bad}21 & ~ & ~ & ~ & ~ & ~ & ~ & 128 & ~ & ~ & ~ \\ 
 \rowcolor{white}
 \multicolumn{2}{r|}{\multirow{-3}{*}{\hspace{-3.1em}\rotatebox[origin=c]{90}{\scriptsize Contracts}}unknown}		& ~ & ~ & ~ &  \cellcolor{good}\np{47518} & ~ & 20 & ~ & ~ & ~ &  \cellcolor{good}\np{45622} & \cellcolor{bad}30 & \cellcolor{good}{248\,238} \\ 
 \rowcolor{white}
 \midrule
\multicolumn{2}{l|}{\# Vulnerabilities}	& 4 & 12+ & 16 & 10 & 7 & 9 & 49 &  \cellcolor{good}57 & 7 & 8 & 5 & 15\\ 
\midrule
\textit{Conkas}		&& ~ & ~ & ~ & ~ & ~ & ~ & ~ & ~ & ~ & & ~ &  \YES\\ 
{ContractFuzzer}	&& ~ & ~ & ~ & ~ & ~ & ~ & ~ & ~ & ~ & \YES & ~ &  \\ 
{Echidna}		&& ~ & ~ & ~ & ~ & ~ & \YES & ~ & ~ & ~ & ~ & ~ &  \\ 
\textit{Ethainter}&& ~ & ~ & ~ & ~ & ~ & ~ & ~ & ~ & ~ &  & ~ &  \YES\\ 
\textit{eThor}		&& ~ & ~ & ~ & ~ & ~ & ~ & ~ & ~ & ~ &  & ~ &  \YES\\ 
ILF			&& ~ & ~ & ~ & ~ & ~ & ~ & ~ & ~ & ~ & \YES & ~ &  \\ 
{HoneyBadger}	&& ~ & ~ & ~ & \YES & ~ & ~ & ~ & ~ & ~ & ~ & \YES &  ~ \\ 
\textit{MadMax}      && ~ & ~ & ~ & ~ & ~ & ~ & ~ & ~ & ~ &  & ~ &  \YES\\ 
Maian		&& ~ & ~ & ~ & \YES & ~ & ~ & \YES & ~ & ~ & ~ & \YES & \YES \\ 
Manticore		&& ~ & ~ & ~ & \YES & \YES & \YES & ~ & ~ & \YES & ~ & \YES &  \\ 
\rowcolor{lightblue}
Mythril		&& ~ & \YES & \YES & \YES & \YES & \YES & \YES & \YES & \YES & \YES & \YES & \YES \\ 
Osiris		&& ~ & ~ & ~ & \YES & ~ & ~ & \YES & ~ & ~ & \YES & \YES & \YES \\ 
\rowcolor{lightblue}
Oyente		&& \YES & \YES & \YES & \YES & \YES & ~ & \YES & ~ & \YES & \YES & \YES & \YES \\ 
\textit{Pakala}      && ~ & ~ & ~ & ~ & ~ & ~ & ~ & ~ & ~ &  & ~ &  \YES\\ 
RA			&& ~ & ~ & ~ & ~ & ~ & ~ & ~ & ~ & ~ & ~ & \YES & \\
Remix-IDE		&& \YES & ~ & \YES & ~ & ~ & ~ & \YES & ~ & ~ & ~ & ~ & \\ 
\rowcolor{lightblue}
Securify		&& \YES & \YES & \YES & \YES & \YES & ~ & \YES & \YES & \YES & \YES & \YES & \YES \\ 
sFuzz		&& ~ & ~ & ~ & ~ & ~ & ~ & ~ & ~ & ~ & \YES & \YES & \\ 
\rowcolor{lightblue}
Slither		&& ~ & ~ & \YES & \YES & \YES & ~ & \YES & \YES & \YES & \YES & \YES & \\ 
\rowcolor{lightblue}
{SmartCheck}	&& \YES & \YES & \YES & \YES & \YES & ~ & \YES & ~ & \YES & \YES & \YES & \\ 
SODA		&& ~ & ~ & ~ & ~ & ~ & ~ & ~ & ~ & ~ & ~ & \YES &  \\
SolHint		&& ~ & ~ & ~ & ~ & ~ & ~ & ~ & ~ & \YES & ~ & \YES &  \\ 
\textit{teEther}     && ~ & ~ & ~ & ~ & ~ & ~ & ~ & ~ & ~ &  & ~ &  \YES\\ 
{SolidityCheck} && ~ & ~ & ~ & ~ & ~ & ~ & \YES & ~ & ~ & ~ & \YES &  \\ 
\textit{Vandal}      && ~ & ~ & ~ & ~ & ~ & ~ & ~ & ~ & ~ &  & ~ &  \YES\\ 
VeriSmart	&& ~ & ~ & ~ & ~ & ~ & ~ & ~ & ~ & \YES & ~ & \YES &  \\ 
VeriSolid	&& ~ & ~ & ~ & ~ & ~ & ~ & ~ & ~ & ~ & ~ & \YES &  \\
\midrule
\multicolumn{2}{l|}{\# Tools}	& 4 & 4 & 6 & 9 & 6 & 3 & 9 & 3 & 8 & 9 & 16 & 12\\
 \bottomrule
     \end{tabular}}
 \end{table}

\subsection{Recent Systematic Reviews on Analysis Tools}
Two studies from early 2022 show that the automated analysis of Ethereum smart contracts has still room for improvement.
\cite{Rameder2022} describe the functionalities and methods of \np{140} tools (\np{83} open source) for automated vulnerability analysis of Ethereum smart contracts.
Their literature review identifies 54 vulnerabilities, with some not addressed by any of the tools.
Moreover, the authors find many tools to be unmaintained. 
\cite{Kushwaha2022tools} provide a systematic review of \np{86} analysis tools with a focus on \np{13} common vulnerabilities.
For quality assessment, they select \np{16} tools, which they test on five vulnerabilities using a ground truth of \np{30} contracts. 

\subsection{Tool Evaluations without Test Sets}
In 2019, two surveys evaluate tools for vulnerability detection by installing them and working through the documentation:
\cite{MdAGS2019tools} investigated \np{27} tools with respect to availability, maturity, methods employed, and security issues detected.
\cite{LopezVivar2020} evaluated \np{18} tools 
regarding the ease of installation, usefulness, and updates.
Both studies do not assess the detection capabilities of the examined tools.

\subsection{Benchmarked Evaluations}
Most closely related to our work are evaluations of tools that actually test them against a set of contracts (benchmark set).
When tool authors compare their own artifact to a few similar and/or popular ones, we consider those works to be intrinsically biased and therefore do not include them.

Among the independent evaluations, we find 11 related works \citep{Dika2017,Parizi2018empirical,Gupta2019,Durieux2020,ghaleb2020effective,Leid2020,Zhang2020a,Dias2021,Ji2021,Ren2021,Kushwaha2022tools}
of which we give an overview in~\autoref{tab:related}.
In the first two rows, we indicate the respective reference and the year when the evaluation was carried out.
Rows three to five list the size of the benchmark set, separated into vulnerable and non-vulnerable contracts, or unknown number of vulnerable contracts.
All references use Solidity files as benchmarks.
Row six indicates the number of different vulnerabilities tested.
We highlight low numbers in red and commendable high numbers in green.
We also list for each tool which evaluation it was part of. 
We highlight the five tools most often used in light blue.
In the last row, there is the total number of tools used in each study.
We highlight the five references using the most tools in mid-blue.

The earliest evaluation was \citep{Dika2017}, which covers four tools tested on five vulnerabilities with a benchmark set of \np{23} vulnerable and \np{21} non-vulnerable contracts.
Regarding the benchmark sets, the number of contracts contained shows a large variety from only \np{10} to almost \np{50000}.
The number of vulnerable contracts in the benchmark set also varies largely from \np{10} to \np{9369}
\footnote{%
It should be noted that for the \textit{wild} benchmark sets, i.e.\ from the contracts actually deployed on the mainchain, the true number of vulnerable contracts and the vulnerabilities they contain is yet unknown.}.
The number of different vulnerabilities varies from \np{4} to \np{57}. 
Several evaluations use their own taxonomy of vulnerabilities. This may be due to the lack of an established taxonomy~\citep{Rameder2022}.

We find a total of \np{20} tools mentioned in the evaluations, while each work selects a subset thereof for its tests.
The number of tools tested varies from three to a maximum of 16. 
The tools most often included in a comparison are Mythril, Oyente, Securify, Slither, and SmartCheck.

 \begin{table}
 \caption{Contributions and Focus of Evaluations with Benchmarks}
 \label{tab:related2}
 \setlength{\tabcolsep}{3pt}
 \centering
     \begin{tabular}{@{}ll|ccccccccccc|r@{}}
 \toprule
& Reference & \SW{\cite{Dika2017}} & \SW{\cite{Parizi2018empirical}} & \SW{\cite{Gupta2020}} & \SW{\cite{Durieux2020}} & \SW{\cite{ghaleb2020effective}} & \SW{\cite{Leid2020}} & \SW{\cite{Zhang2020a}} & \SW{\cite{Dias2021}} & \SW{\cite{Ji2021}} & \SW{\cite{Ren2021}} & \SW{\cite{Kushwaha2022tools}} & \SW{Our study} \\ 
 \midrule
\multirow{5}{*}{\SW{Contribution}}
& systematic review(SLR)& ~ 	& ~ 	& ~ 	& ~ 	& ~ 	& ~ 	& ~ 	& ~ 	& ~ 	& ~ 	& \YES	& ~ 	\\
& weakness classification	& \YES	& ~ 	& \YES 	& ~ 	& ~ 	& ~ 	& \YES	& \YES	& ~ 	& ~ 	& ~ 	& ~ 	\\
& test set of contracts	& ~ 	& ~ 	& \YES 	& \YES	& \YES	& ~ 	& \YES	& ~ 	& ~ 	& ~ 	& ~ 	& \YES	\\
& framework				& ~ 	& ~ 	& ~ 	& \YES	& \YES	& ~ 	& ~ 	& ~ 	& \YES	& ~ 	& ~ 	& \YES	\\
& principles, methods	& ~ 	& ~ 	& ~ 	& ~ 	& \YES 	& ~ 	& ~ 	& ~ 	& ~ 	& \YES	& ~ 	& \YES 	\\
\midrule
\multirow{5}{*}{\SW{Focus}} 
& state-of-the-art		& \YES	& ~ 	& ~ 	& ~ 	& ~ 	& ~ 	& ~ 	& ~ 	& ~ 	& ~ 	& \YES 	& ~ 	\\
& tool effectiveness		& ~ 	& \YES	& \YES	& \YES	& \YES	& \YES	& \YES	& \YES	& \YES	& ~ 	& ~ 	& ~ 	\\
& tool settings			& ~ 	& ~ 	& ~ 	& ~ 	& ~ 	& ~ 	& ~ 	& ~ 	& ~ 	& \YES 	& ~ 	& ~		\\
& creation of test set		& ~ 	& ~ 	& ~ 	& ~ 	& \YES 	& ~ 	& ~ 	& ~ 	& ~ 	& ~ 	& ~ 	& \YES	\\
& timeline				& ~ 	& ~ 	& ~ 	& ~ 	& ~ 	& ~ 	& ~ 	& ~ 	& ~ 	& ~ 	& ~ 	& \YES	\\
\midrule
 \bottomrule
     \end{tabular}
 \end{table}

In~\autoref{tab:related2}, we give an overview of the main contributions and the focus of the benchmarked evaluations.
The contributions include a systematic literature review (\citep{Kushwaha2022tools}), 
their own classification scheme for vulnerabilities or weaknesses (\citep{Dika2017,Gupta2020,Zhang2020a,Dias2021}),
a new or newly assessed benchmark set of contracts (\citep{Gupta2020,Durieux2020,ghaleb2020effective,Zhang2020a}),
a framework for tool execution or test case generation (\citep{Durieux2020,ghaleb2020effective,Ji2021}),
a quantitative tool evaluation (all benchmarked evaluations), or
new principles and methods for tool evaluations.
Regarding principles and methods, \citep{ghaleb2020effective} demonstrate the automated generation of vulnerable contracts by injecting buggy coding patterns, while ~\citep{Ren2021} evaluate settings for tool executions.

As for the focus, most studies address the effectiveness of tools in detecting weaknesses in smart contracts, two studies strive for a general review (\citep{Dika2017,Kushwaha2022tools}), while \citep{Ren2021} aim for insights into the influence of parameter settings onto tool results.

\subsection{Differences to the Benchmarked Studies}
Compared to the benchmarked studies mentioned above, our study stands out in the following aspects.

\ITEM{Focus.} Our study is the only one to focus on the \emph{temporal evolution} of weaknesses and tool behavior as well as on reducing the number of necessary test cases while maintaining \emph{full coverage} of the Ethereum mainchain.
Regarding tool effectiveness, we deliberately do not address it per se -- due to the lack of suitable benchmark sets as the available ones are either small, biased, outdated, or inconsistent \citep{MdAGS2023GT}.
Rather, we are striving for a relative comparison of tools with regard to two aspects: tool behavior over time and against tools that address sufficiently similar weaknesses (via mapping to a common frame of reference).
\ITEM{Input.} We use \emph{runtime bytecode} as input, while the other studies use Solidity source code.
\ITEM{Tools.} We include \emph{further tools} like Conkas, Ethainter, eThor, MadMax, Pakala, teEther, and Vandal. As they accept bytecode only as input, neither of them was used in any of the other studies.
\ITEM{Size.} With a benchmark set of \np{248328} unique contracts from the main chain, we use the \emph{largest number of contracts}.
\ITEM{Contribution.}
Our study features
a \emph{novel method for selecting a benchmark set} that allows for analyzing an entire ecosystem, 
the extension of an \emph{execution framework to work with bytecode as input}, and
the inclusion of \emph{time as a further dimension} to look at weaknesses and reasons for tool behavior.

\subsection{Open Source Frameworks}
For a large-scale evaluation, we need an analysis framework that (i) facilitates the control of multiple tools via a uniform interface, (ii) allows for bulk operation, and (iii) is open source and usable.
SmartBugs~\citep{ferreira2020smartbugs} is such an execution framework released in 2019.
It is still being maintained with 13 contributors and over \np{70} resolved issues.
The framework USCV~\citep{Ji2021} implemented similar ideas in mid-2020.
It comprises an overlapping set of tools and an extension of the ground truth set.
With a total of 10 commits (the latest in mid-2021) and no issues filed, it seems to be neither widely used nor maintained.
Both frameworks target Solidity source code, and thus need to be expanded to work with bytecode.

\newpage
\section{Conclusion}
\label{sec:conclusion}

In this work, we investigated the evolution of smart contract weaknesses as reported by analysis tools.
We managed to cover all 48 million smart contracts deployed on the Ethereum main chain up to block 14\,000\,000, by selecting tools that are able to process runtime bytecode, and by choosing only one representative for each group of contracts with the same skeleton.
In total, we ran 12 tools on \np{248328} contracts with a cumulative execution time of 30 years.
We summarize our contributions and observations.

\ITEM{Skeletons are an effective technique to identify similar contracts.}
Clustering contracts by their skeleton reduces the computing effort as well as the bias that is introduced when sampling contracts from a population that contains some contracts once and others thousandfold.
We show that the validity of studies like ours is not affected by picking only one contract per cluster.

\ITEM{The rate of reported weaknesses decreases over time.}
The tools report a total of \np{1307484} weaknesses, the most common ones being \emph{\SWC{107}} (\percent{184610}{1307484}), \emph{\SWC{104}} (\percent{183277}{1307484}) and \emph{\SWC{101}} (\percent{123255}{1307484}).
The weaknesses are not equally distributed over the study period, though.
By and large, we observe for all tools and all weaknesses a decrease in flagged contracts over time. We offer three explanations.

Some tools are no longer maintained and cannot handle operations added to the EVM later on.
As such operations get more widely used, the tools increasingly fail in their analyses.

Even tools interpreting all operations correctly, may detect weaknesses by code patterns that are tied to specific compiler versions.
As the code generator changes with newer compilers, the code patterns become less effective in indicating the weakness.

But the decrease in flagged contracts can also be observed for maintained and recent tools, which indicates that the weaknesses become indeed less prevalent over time.
This may be attributed to factors like the adoption of good programming practices, public repositories with tested code, enhancements to the programming language Solidity, and checks added by the Solidity compiler.

\ITEM{The analysis tools differ considerably regarding resource consumption and engineering aspects.}
We see large differences in average runtimes, in the number of analyses timing out or running out of memory, and in the number of errors and failures.
These aspects are of relevance in practice, e.g.\ when integrating analysis tools in CI/CD workflows.

\ITEM{The tools agree only partially in their judgment of contracts, with the disagreement increasing over time.}
Our overlap analysis shows that tools targeting the same weakness flag rather different sets of contracts.
The intersection of these sets decreases over time.
We attribute this phenomenon to diverging interpretations of the weaknesses, as precise and commonly accepted definitions are lacking.
Regarding the change over time, our data provides no explanation.

\ITEM{Service to the community.}
In the course of our study, we found several bugs in tools, which we reported either by filing issues or by exchanging emails and engaging in discussions.
The extension of SmartBugs to process bytecode has already been taken up by the framework Centaur\footnote{\url{https://github.com/mchara01/centaur}}.

\ITEM{Recommendations to smart contract developers, tool authors, and the community at large.}
From the experience gathered in this study, we derive the following recommendations and wishes.
\begin{itemize}
\item
  When hardening or auditing smart contracts, use a range of analysis tools, as their approaches and abilities are complementary.
  Grant the tools sufficient resources, memory- and timewise.
\item
  Maintain academic tools for some years after publishing the accompanying article, and keep them public, as a service to the community.
  This allows researchers to evaluate new methods against the state of the art, on recent data.
\item
  Strive for an abstract definition of the weakness addressed, using e.g.\ some formal semantics, execution traces, and path conditions.
  Give a precise definition of the code patterns used to detect the weakness.
  This makes it easier to analyze the scope of tools and to interpret their results.
\item
  Work towards a comprehensive, balanced ground truth.
  Ultimately, many interesting questions regarding the quality of tools and their methods can only be answered by having access to an `oracle' saying true or false.
  This goal is interlinked with the previous one, as the latter determines the meaning of the former.
\end{itemize}

\section*{Data and Code Availability}

The data and the scripts of our study are available from \url{https://figshare.com/s/5efef6335fa98ddc3ae2}.
The dataset of \np{248328} contracts with distinct skeletons has additionally been published at \url{https://github.com/gsalzer/skelcodes}.
SmartBugs is developed as a GitHub project at \url{https://github.com/smartbugs/smartbugs}.
Some utilities for the manipulation of bytecode, like the computation of skeletons, are maintained at \url{https://github.com/gsalzer/ethutils}.

\section*{Acknowledgements}
This project was partially supported by national funds through Funda\c{c}\~ao para a Ci\^encia e a Tecnologia (FCT) under project UIDB/50021/2020.
The project was also partially supported by the CASTOR Software Research Centre.

We are particularly grateful to three anonymous reviewers, whose detailed reports and constructive advice lead to substantial improvements.

\newpage
\bibliographystyle{plainnat}
\bibliography{references}
\appendix
\clearpage
\section*{Appendix}
\begin{center}
\begin{longtable}{l>{\ttfamily}lr}
\caption[Mapping of Tool Findings to SWC Classes]{Mapping of Tool Findings to SWC Classes}
\label{tab:mapping} \\

\toprule
\multicolumn{1}{l}{\textbf{Tool}} & \multicolumn{1}{l}{\textbf{Finding}} & \multicolumn{1}{r}{\hspace{-2cm}\textbf{SWC Class}} \\ 
\midrule
\endfirsthead

\multicolumn{3}{c}%
{{\bfseries \tablename\ \thetable{} -- continued from previous page}} \\
\toprule
\multicolumn{1}{l}{\textbf{Tool}} & \multicolumn{1}{l}{\textbf{Finding}} & \multicolumn{1}{r}{\hspace{-2cm}\textbf{SWC Class}} \\ 
\midrule 
\endhead

\midrule &\multicolumn{2}{r}{{Continued on next page}} \\
\endfoot

\bottomrule
\endlastfoot

Conkas    & Integer\_Overflow                                            & \SwcIntegerOverflowAndUnderflow \\
Conkas    & Integer\_Underflow                                           & \SwcIntegerOverflowAndUnderflow \\
Conkas    & Reentrancy                                                   & \SwcReentrancy \\
Conkas    & Time\_Manipulation                                           & \SwcTimestampDependence \\
Conkas    & Transaction\_Ordering\_Dependence                            & \SwcTxOrderDependence \\
Conkas    & Unchecked\_Low\_Level\_Call                                  & \SwcUncheckedRetVal \\
\midrule
Ethainter & AccessibleSelfdestruct                                       & \SwcUnprotectedSelfdestruct \\
Ethainter & TaintedDelegatecall                                          & \SwcDelegatecallToUntrustedContract \\
Ethainter & TaintedOwnerVariable                                         & \SwcWriteToArbitraryStorage \\
Ethainter & TaintedSelfdestruct                                          & \SwcUnprotectedEtherWithdrawal \\
Ethainter & TaintedStoreIndex                                            & \SwcWriteToArbitraryStorage \\
Ethainter & TaintedValueSend                                             & \SwcUnprotectedEtherWithdrawal \\
\midrule
eThor     & insecure                                                     & \SwcReentrancy \\
\midrule
MadMax    & OverflowLoopIterator                                         & \SwcIntegerOverflowAndUnderflow \\
MadMax    & UnboundedMassOp                                              & \SwcDosWithBlockGasLimit \\
MadMax    & WalletGriefing                                               & \SwcMultipleSends \\
\midrule
Maian     & Destructible                                                 & \SwcUnprotectedSelfdestruct \\
Maian     & Ether\_leak                                                  & \SwcUnprotectedEtherWithdrawal \\
\midrule
Mythril   & Delegatecall\_to\_user\_supplied\_address\_SWC\_112          & \SwcDelegatecallToUntrustedContract \\
Mythril   & Dependence\_on\_predictable\_environment\_variable\_SWC\_116 & \SwcTimestampDependence \\
Mythril   & Dependence\_on\_predictable\_environment\_variable\_SWC\_120 & \SwcWeakRandomness \\
Mythril   & Dependence\_on\_tx\_origin\_SWC\_115                         & \SwcTxOriginUsage \\
Mythril   & Exception\_State\_SWC\_110                                   & \SwcAssertViolation \\
Mythril   & External\_Call\_To\_User\_Supplied\_Address\_SWC\_107        & \SwcReentrancy \\
Mythril   & Integer\_Arithmetic\_Bugs\_SWC\_101                          & \SwcIntegerOverflowAndUnderflow \\
Mythril   & Jump\_to\_an\_arbitrary\_instruction\_SWC\_127               & \SwcArbitraryJump \\
Mythril   & Multiple\_Calls\_in\_a\_Single\_Transaction\_SWC\_113        & \SwcMultipleSends \\
Mythril   & State\_access\_after\_external\_call\_SWC\_107               & \SwcReentrancy \\
Mythril   & Unchecked\_return\_value\_from\_external\_call\_SWC\_104     & \SwcUncheckedRetVal \\
Mythril   & Unprotected\_Ether\_Withdrawal\_SWC\_105                     & \SwcUnprotectedEtherWithdrawal \\
Mythril   & Unprotected\_Selfdestruct\_SWC\_106                          & \SwcUnprotectedSelfdestruct \\
Mythril   & Write\_to\_an\_arbitrary\_storage\_location\_SWC\_124        & \SwcWriteToArbitraryStorage \\
\midrule
Osiris    & Concurrency\_bug                                             & \SwcTxOrderDependence \\
Osiris    & Overflow\_bugs                                               & \SwcIntegerOverflowAndUnderflow \\
Osiris    & Reentrancy\_bug                                              & \SwcReentrancy \\
Osiris    & Timedependency\_bug                                          & \SwcTimestampDependence \\
Osiris    & Underflow\_bugs                                              & \SwcIntegerOverflowAndUnderflow \\
\midrule
Oyente    & Re\_Entrancy\_Vulnerability                                  & \SwcReentrancy \\
Oyente    & Timestamp\_Dependency                                        & \SwcTimestampDependence \\
Oyente    & Transaction\_Ordering\_Dependence\_TOD                       & \SwcTxOrderDependence \\
\midrule
Pakala    & call\_bug                                                    & \SwcUnprotectedEtherWithdrawal \\
Pakala    & delegatecall\_bug                                            & \SwcDelegatecallToUntrustedContract \\
Pakala    & selfdestruct\_bug                                            & \SwcUnprotectedEtherWithdrawal \\
\midrule
Securify  & DAO                                                          & \SwcReentrancy \\
Securify  & DAOConstantGas                                               & \SwcReentrancy \\
Securify  & TODAmount                                                    & \SwcTxOrderDependence \\
Securify  & TODReceiver                                                  & \SwcTxOrderDependence \\
Securify  & TODTransfer                                                  & \SwcTxOrderDependence \\
Securify  & UnhandledException                                           & \SwcUncheckedRetVal \\
Securify  & UnrestrictedEtherFlow                                        & \SwcUnprotectedEtherWithdrawal \\
\midrule
teEther   & Ether\_leak                                                  & \SwcUnprotectedEtherWithdrawal \\
\midrule
Vandal    & Destroyable                                                  & \SwcUnprotectedSelfdestruct \\
Vandal    & OriginUsed                                                   & \SwcTxOriginUsage \\
Vandal    & ReentrantCall                                                & \SwcReentrancy \\
Vandal    & UncheckedCall                                                & \SwcUncheckedRetVal \\
Vandal    & UnsecuredValueSend                                           & \SwcUnprotectedEtherWithdrawal \\
\end{longtable}
\end{center}

\end{document}